\documentclass[gmd, manuscript]{copernicus}

\usepackage[numbered,open,openlevel=2,depth=3]{bookmark}
\hypersetup{pdfpagemode=UseOutlines}
\begin{document}
\nolinenumbers
\title{WRF-JAX v1.1.28: a differentiable regional weather model with fully coupled physics, numerically consistent with WRF and matching 128-process CPU performance on a single GPU}
\Author[1]{Shijin}{Yuan}
\Author[1]{Tianyu}{Shen}
\Author[1][binmu@tongji.edu.cn]{Bin}{Mu}
\Author[1]{Zhen}{Wei}
\Author[1]{Bo}{Qin}
\Author[2]{Guokun}{Dai}
\affil[1]{School of Computer Science and Technology, Tongji University, Shanghai 201804, China}
\affil[2]{Department of Atmospheric and Oceanic Sciences, Fudan University, Shanghai 200438, China}
\equalcontrib{1,2}
\runningtitle{WRF-JAX: a GPU-accelerated differentiable regional weather model}
\runningauthor{S. Yuan et al.}
\received{}
\pubdiscuss{}
\revised{}
\accepted{}
\published{}
\firstpage{1}
\maketitle

\begin{abstract}
Porting a regional weather model in its entirety to a graphics processing unit (GPU) requires complete implementation of the dynamical core and all physical processes, and improved computational efficiency. This paper introduces WRF-JAX v1.1.28, a JAX reimplementation of the WRF 4.5 dynamical core and the complete CONUS physics suite. We conducted 27 ten-day integrations across nine regions and three initialization months, with a horizontal grid spacing of 3\,km and $700\times700$ grid points, and established two numerical-difference baselines. The full-field RMSEs of wind, temperature, and moisture in WRF-JAX and the WRF internal numerical-difference baseline showed similar long-term evolution: the medians of the case-wise RMSE ratios were 1.05--1.07 at 24\,h and 1.02--1.04 at 240\,h. The mean sustained integration time on a single NVIDIA A800 was 175.0\,s per model hour, corresponding to speedups of 1.17 and 1.05 relative to 128-process single-precision WRF runs compiled with gfortran and ifort, respectively.
The experiments further backpropagated through 100 steps with full physics coupling, obtaining all three-dimensional gradient components of the initial UVTQ fields in a single calculation. The full-field Gaussian random-direction evaluation used 500 directions and three perturbation amplitudes per variable; the median relative differences in directional derivatives for zonal wind, meridional wind, dry potential temperature, and water vapour were 2.58--5.22\,\%, 2.94--6.49\,\%, 7.86--10.06\,\%, and 3.51--3.91\,\%, respectively. WRF-JAX combines fully coupled forecasting with differentiation across time steps, providing computable approximate gradients for regional weather simulation and end-to-end optimization, with accuracy comparable to that of the original WRF model and a forward integration speed equivalent to that of the original WRF model running on at least 128 cores.
\end{abstract}

\introduction
Regional numerical weather prediction requires the coupling of rapidly evolving atmospheric dynamics with physical processes that cannot be fully resolved at the grid scale. The Advanced Research version of the Weather Research and Forecasting model (WRF-ARW) provides a widely used computational framework for this purpose~\citep{skamarock2019}. Its scientific value derives not only from the equations of motion themselves but also from the interactions among dynamics, radiation, cloud processes, turbulence, and the land surface. Preserving these interactions is therefore a central requirement when transferring an existing configuration to a different programming and execution environment.

GPU-oriented array-computing frameworks offer another way to express these calculations: they can compile array operations into device kernels and integrate model components into the broader Python ecosystem. However, small differences in initialization formulas, physical states, or the timing of tendency application can affect the subsequent evolution of the model state. Moreover, even when the scientific calculations have been translated correctly, insufficient optimization, such as repeatedly dispatching small device operations, can make integration slow and leave the model impractical. Access to complete model gradients also facilitates optimization tasks such as solving conditional nonlinear optimal perturbation (CNOP) problems, variational data assimilation, and end-to-end training. The traditional WRF adjoint model supports only a subset of simplified physical parameterizations and cannot provide gradients under full physics coupling~\citep{skamarock2019}; in the absence of a usable adjoint model, previous studies have combined dimensionality reduction with gradient-free intelligent algorithms such as particle swarm optimization to solve CNOP problems~\citep{mu2015ppso,zhang2018cnop}. Development must therefore address WRF's numerical accuracy, GPU computational efficiency, and differentiability under full physics coupling together.

Three recent efforts are closely related to this study. PyWRF reconstructs the WRF dynamical core and the WRF single-moment six-class microphysics scheme (WSM6) in PyTorch and demonstrates coupling to a trainable correction module~\citep{chu2026}. That study used a $700\times700\times41$ grid and an 18\,s time step to evaluate eight cases over 6\,h; the model implements only the microphysics parameterization, without radiation, land-surface, boundary-layer, or cumulus parameterizations. The open-source project \texttt{wrf\_gpu} implements WRF-like dynamics, parameterizations, nesting, and compatible input/output in JAX and reports GPU performance for several cases~\citep{wrfgpu2026}. ArWen, released as \texttt{gpuwm}, combines Python, CuPy, and CUDA with physics components derived from WRF~\citep{arwen2026}, and WOOF is a derivative project~\citep{woof2026}. These efforts represent useful attempts at Python weather models, but none has simultaneously met the requirements for numerical consistency, integration speed, and differentiability. They also raise a further question: how can complete coupling between dynamics and physics be preserved while achieving numerical consistency, execution efficiency, and differentiability in long forecasts across multiple regions?

This paper addresses this question through WRF-JAX. WRF-JAX reimplements the WRF 4.5 dynamical core and the radiation, surface exchange, land-surface hydrology, boundary-layer mixing, convection, and microphysics in the CONUS suite as scientific computing functions in JAX, with compilation, column recurrences, and memory management organized for GPUs. Under a common, fully coupled configuration, ten-day integrations across multiple regions and seasons test both sustained operation and numerical consistency.

This study makes four main contributions. First, progressive porting uses methods including runtime coverage, call-dependency analysis, temporary Fortran bridges, and hierarchical difference checks to complete the development of WRF-JAX while retaining fully coupled calculations. Second, a cross-region, cross-season evaluation comprising 27 ten-day integrations is established, and the reference model's own numerical differences are quantified using two methods: comparisons between WRF runs with $8\times16$ and $16\times8$ decompositions, and ensemble consistency tests. The evaluation covers native-grid variables, meteorological pressure-level diagnostics, near-surface fields, and accumulated precipitation, among other outputs. Third, sustained integration speed is quantified from measured logs, with comparisons across CPU platforms, compiler toolchains, and parallel configurations. Fourth, 100-step backpropagation under a fully coupled physics configuration yields the gradients of the initial UVTQ fields, and gradient quality is quantified using independent finite differences in full-field Gaussian random directions. This provides a differentiation interface for CNOP problems, variational data assimilation, and end-to-end optimization, alleviating the limitation that the traditional WRF adjoint model cannot provide gradients under full physics coupling.

\section{WRF-JAX implementation}
\label{sec:implementation}
\subsection{Coverage-guided progressive porting}
\label{sec:development}
The scope of the WRF-JAX port comprises the WRF 4.5 ARW dynamical core and the CONUS physics suite; other, unselected parameterization schemes are outside the scope of this study. Runtime coverage of the selected configuration was analysed using gcov/LCOV~\citep{gccgcov2026}: of the 503\,028 executable source lines counted, 82\,241 were exercised, accounting for 16.3\,\%. Coverage records and source-code analysis were combined to identify the relevant modules, retaining all conditional branches in the selected schemes that depend on the meteorological state. Score-P was used to construct a function call graph~\citep{knuepfer2012}; argument reads and writes and initialization dependencies were then used to organize the porting sequence, progressing from low-level computational functions to physics drivers and time-step scheduling.

During development, the original Fortran subroutines for Thompson, Tiedtke, and RRTMG were first temporarily connected to WRF-JAX through F2PY~\citep{peterson2009}, serving as transitional components that maintained full physics coupling. They were then translated and replaced module by module. Array layouts, data types, initialization, and in-place updates were checked during replacement to ensure consistent state transfer between components. For example, the Thompson column processes were translated and checked for an entire column before horizontal batching was introduced and the write-back to three-dimensional fields was verified. All these components were ultimately translated, and the temporary Fortran bridges were removed.

The port preserves the mixed precision of physical states, constants, and intermediate quantities in the reference implementation, as well as integer conversions for table lookups, recurrence and reduction order, branch thresholds, and the timing of assignments to component outputs.

Numerical verification uses difference comparisons with identical inputs at matching stages. Read-only probes inserted into the original WRF record component inputs, tendencies, and persistent states, and WRF-JAX replays the corresponding calculations to locate the first divergence along the call chain. After local corrections, coupled integrations are repeated to examine error accumulation and feedback. Performance optimizations, including array reformulation, batching, and compilation fusion, follow the same verification procedure, with stage-wise comparisons and complete integrations checking both accuracy and execution time.

Forecast outputs are written sequentially during integration at the configured interval. They include atmospheric and near-surface fields, together with the base-state and perturbation components of pressure and geopotential required for pressure-level diagnostics.

Algorithm~\ref{flow:integration} shows the time-advancement sequence of WRF-JAX using RK3 and the CONUS physics configuration. Physics tendencies are prepared in the first RK stage, and microphysics is executed after all three RK stages; longwave and shortwave radiation processes are called at a separately configured radiation interval.

\begin{algorithm}[ht]
\caption{Time advancement in WRF-JAX}
\label{flow:integration}
\begin{algorithmic}[1]
\REQUIRE Initial fields, static data $\boldsymbol{c}$, boundary data $\boldsymbol{b}(t)$, physics configuration $\mathcal{P}$, time step $\Delta t$, number of steps $N$
\ENSURE Final state $\boldsymbol{x}_N$ and forecast snapshots at specified times
\STATE Initialize model state $\boldsymbol{x}_0$, persistent physics states, radiation clock, and output clock
\FOR{$n=0,\ldots,N-1$}
    \STATE Read boundary data $\boldsymbol{b}_n$ for this step and retain the state at the beginning of the step
    \FOR{RK stage $r=1,2,3$}
        \STATE Prepare stage states and diagnostics, and determine the stage time step and number of acoustic substeps
        \IF{$r=1$}
            \IF{the radiation clock is triggered}
                \STATE Call RRTMG; update and save radiative heating rates and fluxes
            \ENDIF
            \STATE Call the Eta surface layer and Noah land surface in sequence; update surface fluxes and land-surface states
            \STATE Call the MYJ boundary layer; update turbulence states and mixing tendencies
            \STATE Call Tiedtke cumulus; update convective tendencies and precipitation
        \ENDIF
        \STATE Compute dynamical tendencies, incorporate physics tendencies, and apply lateral boundary forcing
        \STATE Perform acoustic substeps to advance momentum, mass, potential temperature, and geopotential
        \STATE Advance water-species and scalar transport; apply positivity constraints and stage boundary conditions
    \ENDFOR
    \STATE Call Thompson microphysics; update thermodynamics, water species, and accumulated precipitation
    \STATE Update diagnostics and boundaries to obtain $\boldsymbol{x}_{n+1}$
    \IF{an output time is reached}
        \STATE Save a forecast snapshot
    \ENDIF
\ENDFOR
\end{algorithmic}
\end{algorithm}

\subsection{GPU acceleration of integration}
\label{sec:execution}
JAX provides array transformations and compilation for Python implementations~\citep{jax2018}. WRF-JAX no longer requires the MPI communication mechanism of the original WRF and can express the dynamical core as domain-wide array operations while retaining the column structure of physics calculations. Backend scheduling handles the compilation and composition of these functions. Recurrences that depend on adjacent vertical levels retain their original order, while independent horizontal columns can be processed in batches.

Compilation is organized into reusable stages, allowing regular time steps and radiation events to reuse their respective compiled computations while controlling compilation size. Static array contents are managed separately from frequently changing states, and a persistent executable cache reduces repeated startup costs for compatible runs. The host coordinates stage calls and output.

Within-column recurrences retain their inter-level dependencies, while independent horizontal columns can execute in parallel or in batches, reducing intermediate arrays and fine-grained device dispatches.

\subsection{Automatic differentiation in WRF-JAX}
\label{sec:ad-implementation}
Explicit states provide an interface for automatic differentiation. Let $\mathcal{F}_N$ denote the composite mapping of $N$ consecutive time steps. Once a scalar objective is defined on the final state, reverse-mode automatic differentiation propagates adjoint variables through the coupled computational chain, obtaining the gradient of the objective with respect to the initial control fields in a single calculation. The reverse calculation reuses the scientific functions of the forecast, with stage checkpointing and recomputation controlling state-storage costs. The Thompson sedimentation loop recomputes inputs and accumulates adjoint variables for the dynamic substeps actually executed, without truncating the loop or stopping state gradients. Branch derivatives correspond to the currently active branch. RRTMG optical-depth table lookups retain the original discrete table values in the forecast, while the reverse calculation uses derivatives of the continuous transmittance and source functions represented by the tables. This rule approximates the continuous radiative response to avoid differentiating the integer indices themselves.

Derivative rules for some low-level operators are expressed as Jacobian--vector products (JVPs), which the automatic differentiation system transposes for backpropagation. For an initial perturbation direction $\boldsymbol{v}$, the forward directional derivative is
\begin{equation}
 \dot{\boldsymbol{x}}_N
 = D\mathcal{F}_N(\boldsymbol{x}_0)\boldsymbol{v}.
 \label{eq:jvp}
\end{equation}
This study uses independent finite differences to check the reverse-mode gradients with respect to the initial UVTQ fields after 100 integration steps~\citep{navon1992}.

Bitcast assignments used to preserve Fortran floating-point semantics require separate derivative handling: forward values are still rounded and stored according to the original rules, whereas derivatives are propagated as real-valued assignments, without differentiating floating-point bit patterns. For quotients spanning a wide range of numerical scales, we use
\begin{equation}
 \dot{y}=\frac{\dot{a}}{b}-\frac{a}{b}\frac{\dot{b}}{b},
 \qquad y=\frac{a}{b},
 \label{eq:stable-quotient-jvp}
\end{equation}
to avoid explicitly forming the intermediate results $b^2$ or $a\dot{b}$. This rule is used for divisions involving extreme scales in radiative optical quantities, convective evaporation, and moisture-transport limiters; unselected expressions in masked branches also avoid invalid division.

\section{Evaluation design}
\label{sec:methods}
\subsection{Evaluation regions and periods}
\begin{figure}[ht]
\centering
\includegraphics[width=\textwidth]{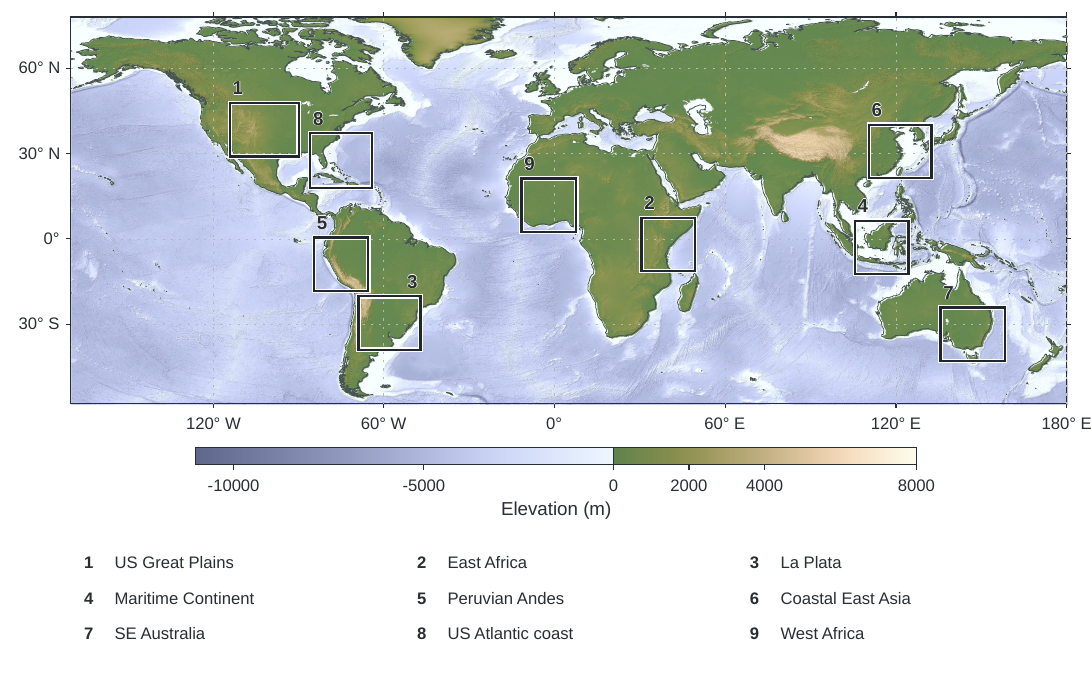}
\caption{The nine regions covered by the 27-case evaluation in this study. Dark-grey outlines indicate the 3\,km Mercator simulation domains, shown in longitude and latitude on the global overview map. Numbers correspond to Table~\ref{tab:regions}. The background is the 2-arcmin terrain elevation from SRTM15+~\citep{tozer2019}, displayed using the oleron colour map with terrain shading~\citep{crameri2023}.}
\label{fig:domains}
\end{figure}

The experiments comprise nine regions, each initialized at 00:00\,UTC on 1 March, 1 July, and 1 November 2021 and integrated for ten days. The regions cover midlatitude continents, tropical land and ocean, coastal environments, and pronounced terrain gradients (Fig.~\ref{fig:domains} and Table~\ref{tab:regions}). With a common resolution and physics configuration, these regions provide different physical states; the same months also represent different seasonal phases in the Northern and Southern Hemispheres.

\begin{table}[ht]
\caption{Centre coordinates and sampling rationale for the nine regions. Each region includes ten-day integrations initialized in three months.}
\label{tab:regions}
\centering
\begin{tabularx}{\textwidth}{rp{2.7cm}p{3.5cm}X}
\tophline
No. & Region & Centre coordinates & Sampling rationale \\
\middlehline
1 & US Great Plains & $39.00^{\circ}$\,N, $102.00^{\circ}$\,W & Continental interior, convection, and western terrain gradients \\
2 & East Africa & $2.00^{\circ}$\,S, $40.00^{\circ}$\,E & Tropical highlands and the western Indian Ocean margin \\
3 & La Plata & $30.00^{\circ}$\,S, $58.00^{\circ}$\,W & Southern Hemisphere continental and coastal environments \\
4 & Maritime Continent & $3.00^{\circ}$\,S, $115.00^{\circ}$\,E & Humid tropical convection and heterogeneous land--sea distribution \\
5 & Peruvian Andes & $9.00^{\circ}$\,S, $75.00^{\circ}$\,W & Steep terrain and strong coastal--inland contrasts \\
6 & East Asian coast & $31.23^{\circ}$\,N, $121.47^{\circ}$\,E & East Asian coastal environments and seasonal contrasts \\
7 & Southeastern Australia & $34.00^{\circ}$\,S, $147.00^{\circ}$\,E & Southern Hemisphere midlatitude land, coast, and ocean \\
8 & US Atlantic coast & $28.00^{\circ}$\,N, $75.00^{\circ}$\,W & Subtropical western Atlantic and adjacent land \\
9 & West Africa & $12.00^{\circ}$\,N, $2.00^{\circ}$\,W & Tropical continental moisture and seasonal precipitation gradients \\
\bottomhline
\end{tabularx}
\end{table}

The regions were selected to cover contrasting environments. This study selected 27 cases with complete ten-day Fortran reference results and conducted the corresponding WRF-JAX integrations and baseline comparisons.

\subsection{WRF-JAX physics configuration}
\label{sec:physics}
The evaluation uses the CONUS physics suite, comprising Thompson microphysics~\citep{thompson2008}, RRTMG longwave and shortwave radiation~\citep{iacono2008}, Mellor--Yamada--Janjic boundary-layer mixing~\citep{janjic2001}, the corresponding Eta surface-layer scheme, the Noah land-surface model~\citep{chen2001}, and the modified Tiedtke cumulus scheme~\citep{zhang2011}. All these physical processes remain enabled in both accuracy and performance tests, and radiation calculation time is included in sustained integration timing.

Table~\ref{tab:configuration} lists the configuration used for the 27 ten-day experiments. The WRF-JAX simulation domain, initialization time, horizontal resolution, grid size, integration time step, integration duration, and output interval can be adjusted to the needs of each case. WRF-JAX and the reference WRF use identical settings for each case.

\begin{table}[ht]
\caption{Configuration used for the 27 ten-day experiments. The horizontal grid-point counts include staggered-grid endpoints; three-dimensional mass-point fields in the output files have dimensions of $699\times699\times44$.}
\label{tab:configuration}
\centering
\begin{tabular}{ll}
\tophline
Item & Configuration \\
\middlehline
Reference model and preprocessing system & WRF 4.5 and WPS 4.5 \\
Projection and horizontal grid spacing & Mercator; 3\,km in both directions \\
Horizontal grid-point counts & $700\times700$ \\
Vertical levels & 44 mass levels (45 vertical interfaces) \\
Integration time step and duration & 18\,s; 240\,h (48\,000 steps) \\
Lateral boundary data interval & 6\,h \\
History output interval & 3\,h \\
Model top & 50\,hPa \\
Microphysics & Thompson \\
Longwave and shortwave radiation & RRTMG \\
Radiation call interval & 10\,min \\
Planetary boundary layer & Mellor--Yamada--Janjic \\
Surface layer & Eta \\
Land surface & Noah \\
Cumulus parameterization & Modified Tiedtke \\
Lateral boundary width & 5 grid points \\
\bottomhline
\end{tabular}
\end{table}

Cumulus parameterization on the 3\,km grid retains the modified Tiedtke setting of the reference model to keep the physics configurations of the two implementations consistent.

\subsection{Driving data}
Initial and lateral boundary conditions are taken from the 0.25$^{\circ}$ NCEP Global Data Assimilation System/Final Analysis (GDAS/FNL) data, sampled every 6\,h~\citep{ncep2015}. WPS 4.5 and the WRF real-data preprocessing program generate the initial and boundary fields. Both implementations use the same preprocessed files for a given case, with lateral boundaries updated throughout the ten-day integration.

Each WRF-JAX forecast uses one NVIDIA A800-SXM4-80GB and four CPU threads. The JAX version is 0.9.2.

The Fortran WRF reference consists of the 128-process runs with the default $8\times16$ decomposition (\texttt{nproc\_x=8}, \texttt{nproc\_y=16}) for all 27 cases. Each case was also integrated for the same ten-day period with a $16\times8$ decomposition using the same number of processes, forming a paired WRF internal-difference baseline. Initial and boundary conditions, grid, time step, and physics configuration were identical between the two runs.

Each WRF-JAX case produces hourly history output files. This study selects eight forecast lead times---3, 6, 12, 24, 72, 120, 168, and 240\,h---and computes full-field statistics separately at each time to characterize short-term differences and long-term evolution.

\subsection{Methods for evaluating numerical consistency}
\label{sec:metrics}
\subsubsection{Error metrics}
For a variable $v$ at a fixed forecast lead time $t$, let $x_i$ and $r_i$ denote the WRF-JAX and reference values at the same grid point, respectively. Root-mean-square error (RMSE), mean absolute error (MAE), and normalized root-mean-square error (nRMSE) are calculated independently for each case:
\begin{align}
 \mathrm{RMSE}_{v,t} &= \left[\frac{1}{N}\sum_{i=1}^{N}(x_i-r_i)^2\right]^{1/2}, \label{eq:rmse}\\
 \mathrm{MAE}_{v,t} &= \frac{1}{N}\sum_{i=1}^{N}|x_i-r_i|, \label{eq:mae}\\
 \mathrm{nRMSE}_{v,t} &=100\frac{\mathrm{RMSE}_{v,t}}{\left[N^{-1}\sum_{i=1}^{N}r_i^2\right]^{1/2}}\,\%. \label{eq:nrmse}
\end{align}
Equation~(\ref{eq:nrmse}) normalizes by the root-mean-square amplitude of the reference field. If the reference amplitude is zero, nRMSE is undefined; the corresponding absolute errors are still included in the statistics, and normalized summaries record the number of cases with valid denominators. This treatment mainly concerns nighttime shortwave fluxes.

The evaluation assigns equal weight to all native grid points, including the lateral boundary regions. Three-dimensional winds are compared on their native staggered grids, without vertical or horizontal averaging before error calculation. The variable \texttt{T} is perturbation potential temperature, $\theta'=\theta-300$\,K. Both \texttt{QVAPOR} and \texttt{QRAIN} are mixing ratios. Near-surface diagnostics include 2\,m temperature and humidity, 10\,m wind components, and surface pressure; \texttt{RAINNC} and \texttt{RAINC} denote accumulated explicit and convective precipitation, respectively. Downward shortwave and longwave fluxes at the surface are also evaluated.

Cross-case means, medians, and interquartile ranges are summarized from the metrics calculated independently for each case, with equal weight assigned to every case. All statistical metrics are computed for instantaneous values at the specified forecast times.

Evaluation of common diagnostics provides another meteorological perspective. Total pressure is obtained by adding the base-state and perturbation components; total geopotential is converted to height using $g=9.81$\,m\,s$^{-2}$ and averaged onto mass levels. Air temperature is calculated as $\theta\left[p/(10^5\,\mathrm{Pa})\right]^{R_d/c_p}$, where $R_d=287.0$\,J\,kg$^{-1}$\,K$^{-1}$ and $c_p=1004.5$\,J\,kg$^{-1}$\,K$^{-1}$. Winds are first destaggered and then rotated into geographic components. Linear interpolation in log-pressure coordinates is subsequently performed to 500 and 850\,hPa. Points without valid bracketing levels, including pressure surfaces below the terrain, retain a missing-data mask. Pressure-level RMSE and MAE use all columns valid in both models.

The Supplement presents seven representative outputs for precipitation, radiation, and surface temperature for all 27 cases, together with pressure-level diagnostics at 24 and 240\,h. RAINNC and RAINC use accumulated grid-scale and cumulus precipitation, respectively, from initialization to the specified time; the other variables use the output states or diagnostics at that time. Figures are organized by initialization month, with the nine regions as rows and two groups of three columns placed side by side to compare WRF, WRF-JAX, and their differences at the two lead times.

\subsubsection{WRF internal-difference baseline}
\label{sec:baseline}
Finite-precision rounding is an unavoidable source of error in numerical weather prediction calculations. Floating-point arithmetic is not associative, and changing the order of operations can change the results~\citep{lang2021}. Compiler optimizations and parallel decomposition affect the order of expression evaluation and reduction accumulation, and the resulting small differences can propagate through nonlinear integration. The official WRF guide to bitwise reproducibility checks requires results obtained with different process counts to be checked with compiler optimization disabled, and explicitly notes that the combination of compiler optimization and parallelization may affect simulations so that results are no longer bitwise identical~\citep{wrfbitwise2020}.

Operational forecasting already has precedents for accepting such numerical differences following forecast-quality testing. Since IFS Cycle 47r2, ECMWF has used single-precision computation for operational deterministic and ensemble forecasts. Single- and double-precision integrations are not bitwise identical, but systematic evaluation showed that forecast skill was retained at reduced precision, freeing computational resources to increase the vertical resolution of the ensemble forecasts~\citep{lang2021}. This demonstrates that numerical implementations of operational models can tolerate finite-precision differences while maintaining forecast quality.

To establish a reference level for WRF-JAX differences, this study compared original WRF results obtained with gfortran and ifort compiler configurations, single and double precision, and multiple parallel decompositions, and used a variant of the ensemble consistency test (ECT). This variant draws on the approach of constructing a reference ensemble through small perturbations to the initial fields~\citep{baker2015}, using full-field RMSE and MAE to quantify internal ensemble differences and the differences between WRF-JAX and the reference ensemble.

The ensemble experiments cover nine regions and three initialization months, giving 27 cases. For each case, independent random single-precision floating-point perturbations of 1\,ULP (unit in the last place) are applied at every three-dimensional grid point of the initial dry potential temperature field to generate 10 WRF reference members, each integrated for 6\,h with hourly output. RMSE is calculated using all native grid points for $U$, $V$, dry potential temperature, and water vapour mixing ratio. Within each case, pairwise comparisons among the 10 WRF members yield 45 RMSE values, whose arithmetic mean serves as the WRF internal ensemble-difference baseline; comparisons of WRF-JAX with the 10 members yield 10 RMSE values, whose arithmetic mean characterizes the difference between WRF-JAX and the reference ensemble. For each variable and forecast lead time, the two types of case-mean RMSE are then averaged arithmetically over the 27 cases, and the percentage increase in the WRF-JAX mean RMSE relative to the mean internal ensemble RMSE is calculated. Full-field MAE is also calculated for $U$, $\theta'$, and $q_r$, using the same member pairing and within-case arithmetic averaging to obtain the ensemble comparisons in Fig.~\ref{fig:pywrf}.

The parallel-decomposition difference baseline arises from sparse numerical perturbations introduced by changes in parallel decomposition and operation order, whereas the internal ensemble-difference baseline arises from non-sparse numerical perturbations applied at all three-dimensional grid points of the initial dry potential temperature field; in a chaotic system, however, the two rapidly converge. For the common variables and forecast lead times tested, the relative differences among the RMSE baselines obtained by changing compiler configurations, floating-point precision, parallel decomposition, and using ensemble experiments were all within 1\,\%. This held both for the case-wise RMSE baselines and for the RMSE baselines averaged over the 27 cases.

Considering the computational cost of extending the ten-member ensemble to ten-day integrations for all 27 cases, the ten-day evaluation ultimately uses the $8\times16$ and $16\times8$ decomposition pair to quantify WRF internal differences consistently. The six-hour short-term evaluation against existing work uses both the WRF internal ensemble differences and the differences between WRF runs with $8\times16$ and $16\times8$ decompositions as baselines, and the results show close agreement between the two baselines.

The parallel-decomposition baseline comparison keeps initial and boundary conditions, grid, time step, and physics configuration identical. Using the WRF run with an $8\times16$ decomposition as the common reference, errors are calculated for WRF-JAX and the WRF run with a $16\times8$ decomposition relative to this reference, with the latter referred to as the parallel-decomposition difference baseline. Both comparisons use the same native grids, 14 variables, and Eqs.~(\ref{eq:rmse})--(\ref{eq:nrmse}), with nRMSE normalized by the root-mean-square amplitude of the reference field from the $8\times16$ decomposition. For each variable and forecast lead time, the ratio of WRF-JAX RMSE to baseline RMSE is first calculated for each case, and its cross-case median is then summarized; a ratio close to 1 indicates similar magnitudes of the two types of differences. The ratio is recorded as undefined when baseline RMSE is zero. This paired metric compares WRF-JAX's numerical differences with the original WRF's internal differences.

\subsection{Methods for evaluating integration efficiency}
\label{sec:timing}
Performance is evaluated using sustained integration time, defined as the wall-clock seconds required for integration divided by the corresponding number of model hours, excluding initialization, compilation, and output costs. The ten-day experiments discard the first output interval, which includes startup, and both WRF-JAX and Fortran WRF are evaluated over the matching 3--240\,h interval. WRF-JAX timing likewise excludes file input/output time.

The original WRF logs contain both time-step and disk-write timings, and the time-step timings include disk-write time. Therefore, the main time-step durations are summed and the disk-write time associated with those time steps is subtracted:
\begin{equation}
 C=\frac{\sum_{k\in I}\tau_{k,\mathrm{main}}-\sum_{k\in I}\tau_{k,\mathrm{write}}}{(t_1-t_0)/3600},
 \label{eq:timing}
\end{equation}
where $C$ is measured in seconds per model hour, and $I$ denotes the matching model-time interval. The overall speedup is the mean WRF time divided by the mean WRF-JAX time for the same set of paired cases; a value greater than one indicates faster GPU integration.

\subsection{Validation of WRF-JAX differentiability}
\label{sec:ad-methods}
The differentiability validation experiment uses a 3\,km case initialized at 12:00\,UTC on 27 June 2021, also with an 18\,s time step. The model is integrated for 100 steps (30\,min) on a single A800, retaining dynamics, lateral boundaries, and the complete CONUS physics suite. The radiation time step is 10\,min; in the actual run, radiation tendencies are updated at steps 1, 34, 67, and 100.

The initial control variables are zonal wind, meridional wind, dry potential temperature perturbation, and water vapour mixing ratio, denoted by $\boldsymbol{x}_0=(U_0,V_0,\theta'_0,q_{v,0})$ and abbreviated as UVTQ, where $\theta'=\theta-300$\,K.

The scalar objective for backpropagation is the mean internal moist potential temperature perturbation in the final state at step $N$:
\begin{equation}
 L(\boldsymbol{x}_0)=\frac{1}{|\Omega|}
 \sum_{(i,k,j)\in\Omega}\theta'_m(N\Delta t;i,k,j),
 \label{eq:gradient-loss}
\end{equation}
where $N=100$ and $\Omega$ comprises all 44 mass levels. The function \texttt{jax.value\_and\_grad} is called to compute $\boldsymbol{g}=\nabla_{\boldsymbol{x}_0}L$.

Each direction acts on only one control field and covers all its three-dimensional grid points. Independent standard Gaussian random numbers $z_i\sim\mathcal{N}(0,1)$ are generated and normalized by $s=(n_f^{-1}\sum_i z_i^2)^{1/2}$, where $n_f$ is the dimensionality. For $U$, $V$, and $T$, we set $v_i=z_i/s$ and calculate the central finite difference~\citep{jaxautodiff2026}
\begin{equation}
 D_{\mathrm{FD}}(h)=
 \frac{L(\boldsymbol{x}_0+h\boldsymbol{v})-
       L(\boldsymbol{x}_0-h\boldsymbol{v})}{2h}.
 \label{eq:reverse-fd}
\end{equation}
For water vapour, non-negative Gaussian directions $v_i=|z_i|/s$ are used, with the same absolute amplitude scale throughout the field to avoid producing negative values, and the one-sided finite difference is calculated as
\begin{equation}
 D_{\mathrm{FD}}^{+}(h)=
 \frac{L(\boldsymbol{x}_0+h\boldsymbol{v})-L(\boldsymbol{x}_0)}{h}.
 \label{eq:reverse-fd-q}
\end{equation}
The amplitude $h$ has the same units as the perturbed variable. With the direction components for the other three control fields set to zero, the directional projection of the reverse-mode gradient is calculated as
\begin{equation}
 D_{\mathrm{AD}}=\boldsymbol{g}^{\mathsf T}\boldsymbol{v}
 =\sum_{i=1}^{n}g_i v_i,
 \label{eq:gradient-adjoint}
\end{equation}
where $n$ is the total dimensionality of the three-dimensional fields of all four UVTQ variables. The check compares
\begin{equation}
 \boxed{D_{\mathrm{FD}}(h)\approx D_{\mathrm{AD}}},
 \label{eq:gradient-check}
\end{equation}
with $D_{\mathrm{FD}}^{+}$ used for water vapour. The relative difference is defined as
\begin{equation}
 r(h)=\frac{|D_{\mathrm{AD}}-D_{\mathrm{FD}}(h)|}{|D_{\mathrm{FD}}(h)|}.
 \label{eq:gradient-relative-error}
\end{equation}
Finite differences require two perturbed integrations for each wind or dry potential temperature direction at each amplitude; water vapour requires one perturbed integration and reuses the baseline objective value.

The amplitudes are $h=0.3,0.1,0.03$\,m\,s$^{-1}$ for $U$ and $V$, $0.1,0.03,0.01$\,K for $T$, and $1\times10^{-5},3\times10^{-6},1\times10^{-6}$\,kg\,kg$^{-1}$ for $Q$. The results report the 25th percentile, median, 75th percentile, and arithmetic mean of the relative differences in directional derivatives.

\section{Evaluation of the WRF-JAX model}
\label{sec:evaluation}
\subsection{Numerical consistency evaluation}
\label{sec:results}
\subsubsection{Numerical consistency evaluation of ten-day integrations}
\begin{figure}[ht]
\centering
\includegraphics[width=\textwidth]{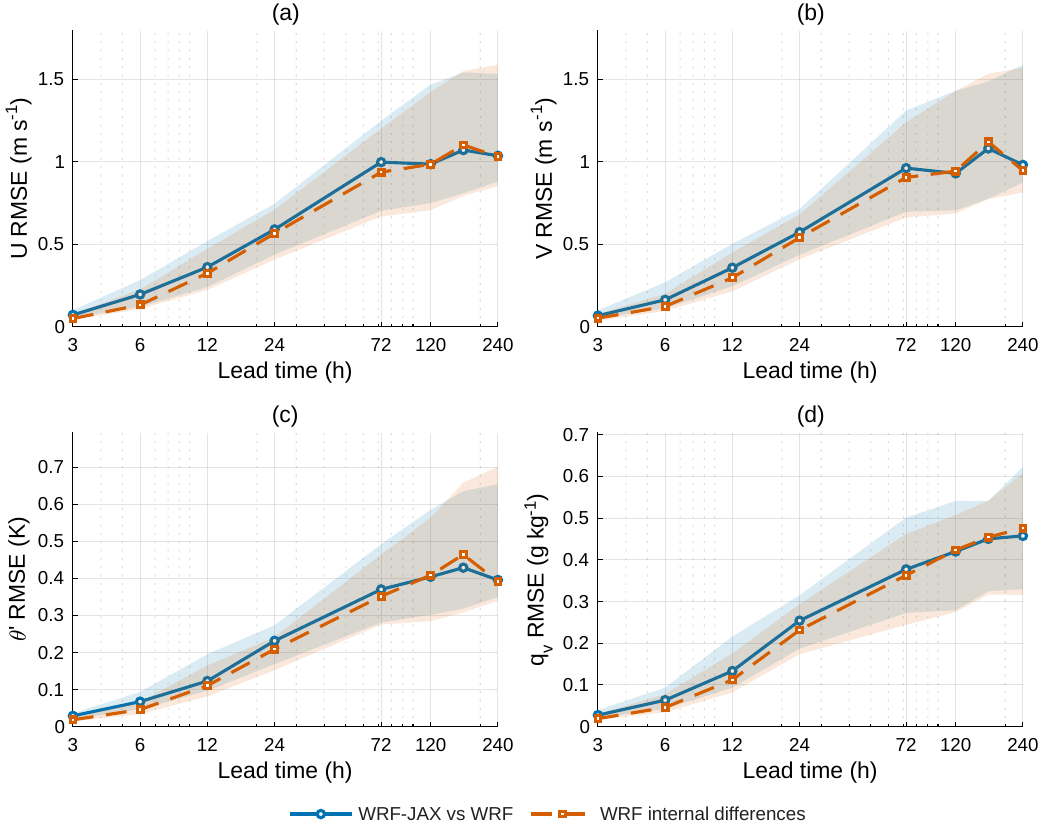}
\caption{Evolution of full-field RMSE for the 27 cases: (a) $U$; (b) $V$; (c) $\theta'$; (d) water vapour. Blue lines denote WRF-JAX vs WRF, and orange lines denote WRF internal differences, calculated as the RMSEs of WRF-JAX and WRF with a $16\times8$ decomposition, respectively, relative to the same WRF reference with an $8\times16$ decomposition. Both lines are cross-case medians, with shading of the same colour spanning the 25th--75th percentiles; symbols mark the eight analysis times with measured output. The horizontal axis is logarithmic, and water vapour is expressed in g\,kg$^{-1}$.}
\label{fig:lead}
\end{figure}

Figure~\ref{fig:lead} shows that errors increase with lead time and exhibit pronounced regional dependence. At 24\,h, the median RMSEs of $U$ and $V$ are 0.5888 and 0.5722\,m\,s$^{-1}$, respectively, while those of $\theta'$ and water vapour mixing ratio are 0.2315\,K and 0.2540\,g\,kg$^{-1}$. By 240\,h, the median wind-component RMSEs increase to 1.0351 and 0.9798\,m\,s$^{-1}$, with $\theta'$ at 0.3957\,K and water vapour at 0.4573\,g\,kg$^{-1}$. From day one to day ten, the medians of the four variables increase to approximately 1.7--1.8 times their initial values, exhibiting clearly nonlinear growth.

\begin{figure}[ht]
\centering
\includegraphics[width=\textwidth]{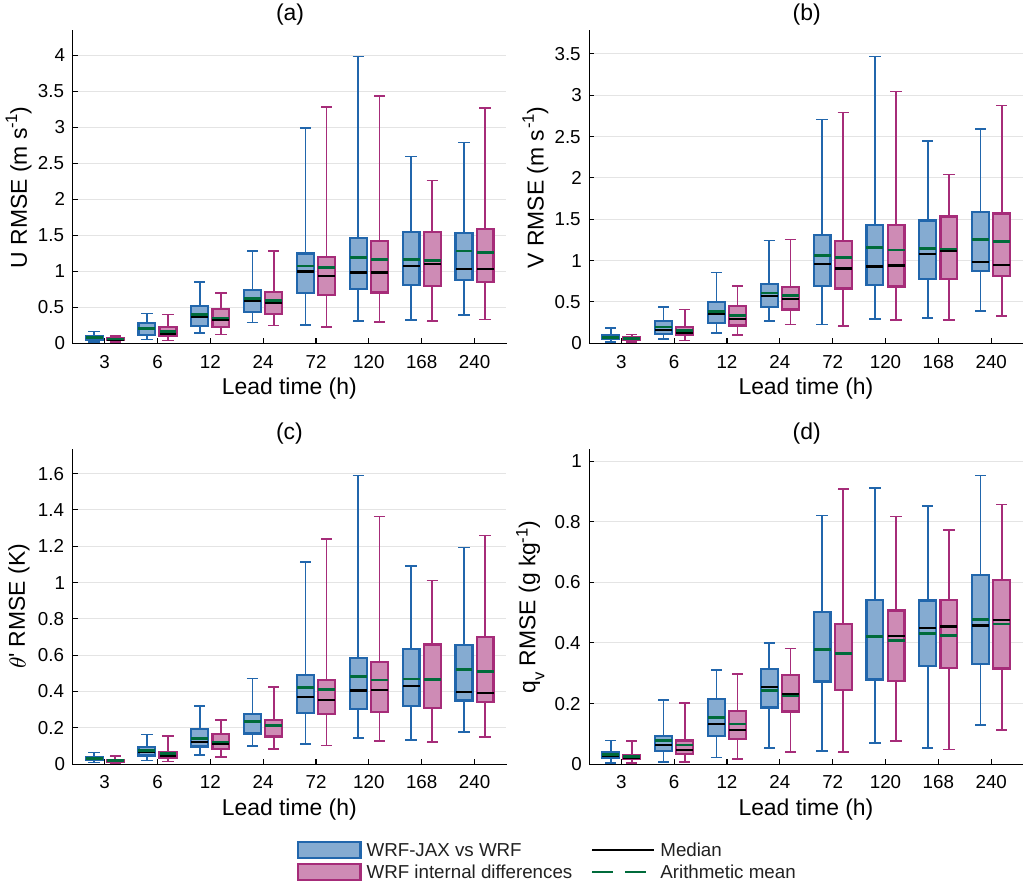}
\caption{Cross-case RMSE distributions at eight forecast lead times, with variables in the same order as in Fig.~\ref{fig:lead}. At each lead time, the dark-blue box on the left denotes WRF-JAX vs WRF, and the magenta box on the right denotes WRF internal differences. These are calculated as the RMSEs of WRF-JAX and WRF with a $16\times8$ decomposition, respectively, relative to the same WRF reference with an $8\times16$ decomposition, with 27 cases in each box. Boxes span the 25th--75th percentiles, black horizontal lines indicate medians, short green dashed lines indicate arithmetic means, and whiskers extend to the minimum and maximum. The eight analysis times are equally spaced along the horizontal axis, and water vapour is expressed in g\,kg$^{-1}$.}
\label{fig:lead-box}
\end{figure}

Figure~\ref{fig:lead-box} complements these results with box plots of the cross-case distributions at each lead time, showing means, medians, and extremes together. The boxes and whiskers of the two comparisons evolve similarly with lead time. At longer lead times, the means of wind and potential temperature exceed their medians, reflecting the influence of a few high-error cases on the averages. The corresponding MAE and nRMSE evolution is shown in Appendix Figs.~\ref{fig:lead-mae} and \ref{fig:lead-nrmse}.

In Fig.~\ref{fig:lead}, both the WRF error baseline and the WRF-JAX errors increase with lead time, and their cross-case interquartile ranges overlap extensively. The two become similar at longer lead times. At 24\,h, the median RMSEs of the two wind components in the WRF internal numerical-difference baseline are 0.5651 and 0.5392\,m\,s$^{-1}$, compared with 0.5888 and 0.5722\,m\,s$^{-1}$ for WRF-JAX; the medians of the case-wise RMSE ratios for the four variables are 1.050--1.066. By 240\,h, the median baseline RMSEs of $U$, $V$, $\theta'$, and water vapour are 1.0304\,m\,s$^{-1}$, 0.9456\,m\,s$^{-1}$, 0.3913\,K, and 0.4760\,g\,kg$^{-1}$, respectively; the medians of the case-wise ratios are 1.029, 1.032, 1.024, and 1.038, respectively. Figure~\ref{fig:mpi} retains the case-wise pairing and shows the consistency between regional differences and the cross-case summaries.

This comparison also extends to near-surface, precipitation, and radiation variables (Tables~\ref{tab:rmse}--\ref{tab:nrmse}). At 240\,h, the median RAINNC RMSEs for WRF-JAX and the WRF internal numerical-difference baseline are 13.46 and 12.63\,mm, respectively, with a median case-wise ratio of 1.091; the corresponding ratio is 0.990 for rainwater mixing ratio, 1.035 for 2\,m temperature, and 1.014 for surface downward longwave flux. Thus, the larger absolute differences over long periods also occur among the original WRF's own numerical results.

\begin{figure}[htp]
\centering
\includegraphics[width=\textwidth]{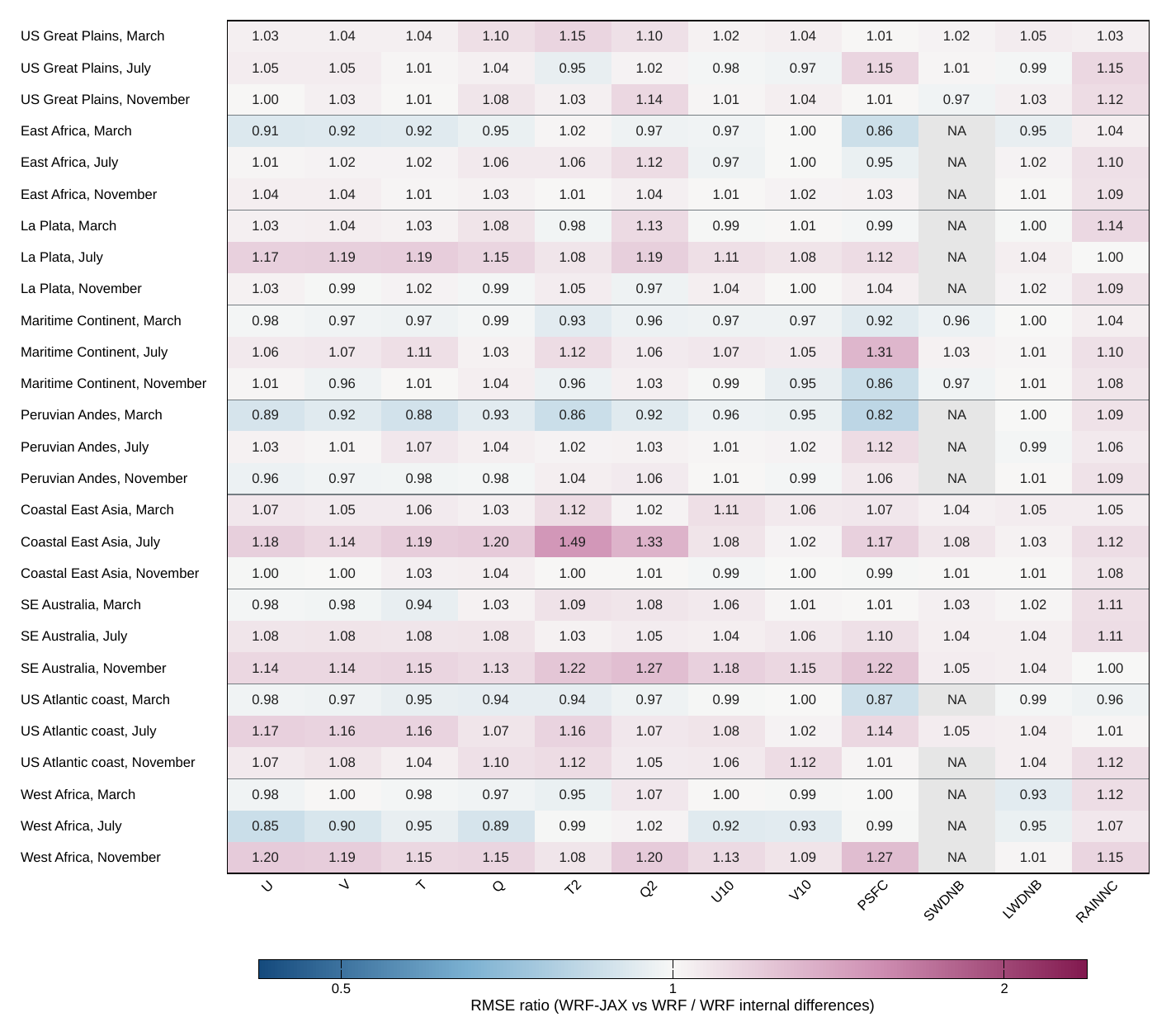}
\caption{Case-wise RMSE ratios at 240\,h. Each cell gives the RMSE between WRF-JAX and WRF divided by the RMSE between WRF runs using $8\times16$ and $16\times8$ decompositions. A ratio of 1 indicates equal magnitudes of the two sets of differences; the colour scale uses a logarithmic mapping centred on 1, and numbers show the actual ratios. Horizontal lines separate the nine regions, each containing three cases initialized in March, July, and November. $T$ is dry potential temperature perturbation and $Q$ is water vapour mixing ratio; T2 and Q2 are 2\,m temperature and water vapour, U10 and V10 are 10\,m wind components, PSFC is surface pressure, and SWDNB and LWDNB are surface downward shortwave and longwave fluxes. RAINNC is grid-scale precipitation accumulated from initialization to 240\,h; all other fields are instantaneous at that time. Grey NA denotes zero nighttime downward shortwave radiation flux, for which the ratio is undefined.}
\label{fig:case-overview}
\end{figure}

Figure~\ref{fig:case-overview} organizes 12 variables at 240\,h into a case-wise matrix. The three-dimensional wind, temperature, and moisture RMSE ratios for the 27 cases range from 0.85 to 1.20.

\subsubsection{Numerical consistency evaluation of six-hour short-term integrations}
\begin{figure}[ht]
\centering
\includegraphics[width=\textwidth]{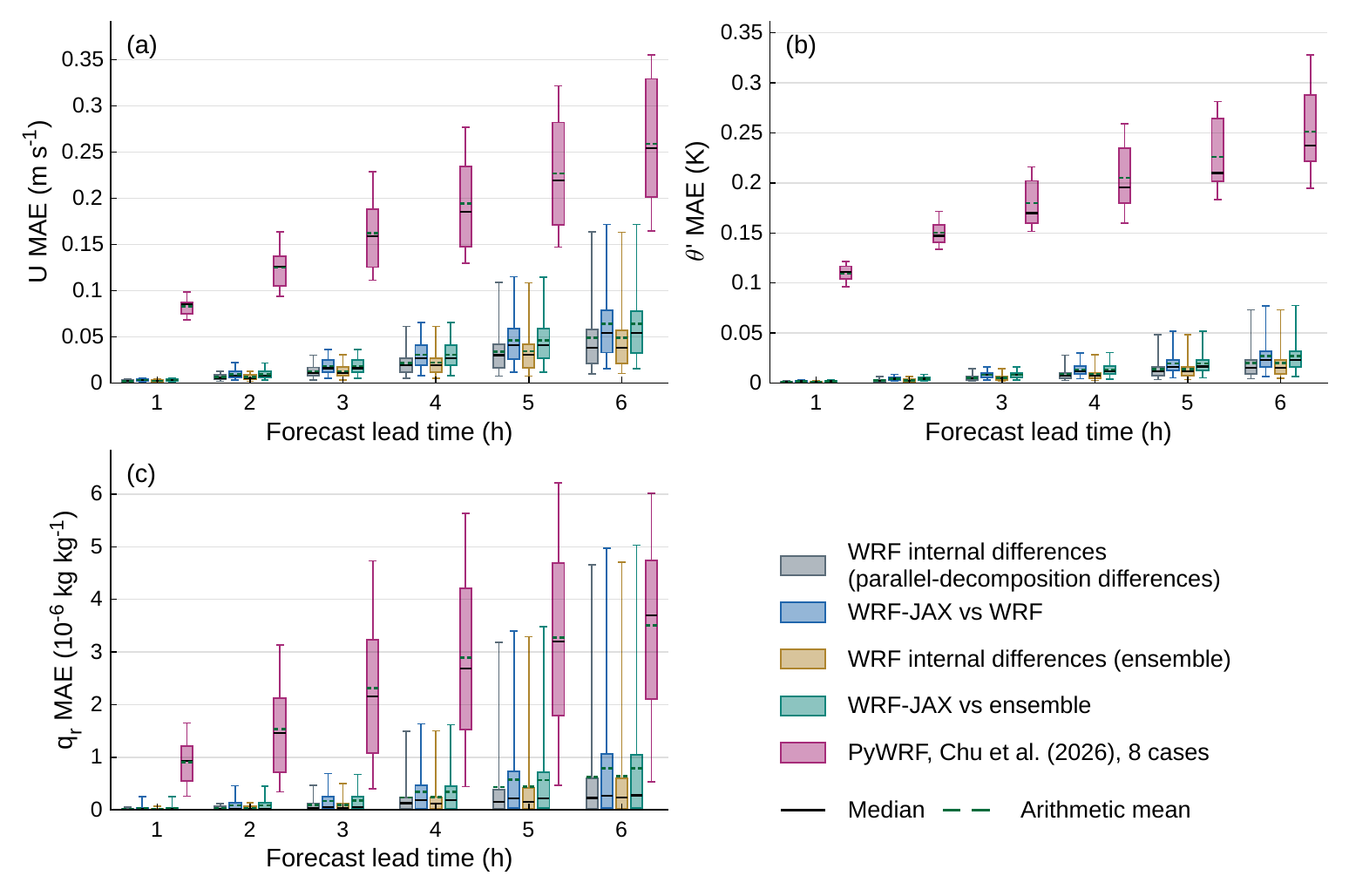}
\caption{Cross-case MAE box-plot comparisons at 1--6\,h: (a) zonal wind; (b) perturbation potential temperature; (c) rainwater mixing ratio. At each time, from left to right, the grey box denotes WRF internal differences, blue denotes WRF-JAX vs WRF, yellow denotes WRF internal ensemble differences, cyan denotes WRF-JAX vs ensemble, and magenta denotes the PyWRF results from Fig.~3 of \citet{chu2026}. The first two groups give the MAEs of WRF with a $16\times8$ decomposition and WRF-JAX, respectively, relative to the same WRF reference with an $8\times16$ decomposition. For the two ensemble comparisons, each case statistic is, respectively, the arithmetic mean of 45 MAEs from pairwise comparisons among 10 WRF members, and the arithmetic mean of 10 MAEs from comparisons between WRF-JAX and those members. Each of the first four groups contains 27 cases and uses measured hourly output from six-hour experiments; PyWRF contains the eight cases described in its paper. Boxes span the 25th--75th percentiles, black horizontal lines indicate medians, short green dashed lines indicate arithmetic means, and whiskers extend from the minimum to the maximum.}
\label{fig:pywrf}
\end{figure}

Following the cross-case box-plot representation in Fig.~\ref{fig:lead-box}, Fig.~\ref{fig:pywrf} presents the six-hour evaluation results of WRF-JAX alongside those of an existing reimplementation, using the statistical approach in Fig.~3 of \citet{chu2026}. The variables compared are zonal wind $U$, perturbation potential temperature $T=\theta'$, and rainwater mixing ratio $q_r$ (QR in the PyWRF figure and QRAIN here). Each statistical sample corresponds to the MAE of one case at a specified forecast time; for ensemble comparisons, it is the arithmetic mean of the member-pair MAEs within that case. Quartiles, means, and extremes are then calculated across the case set for each group.

At all six measured lead times from 1 to 6\,h, WRF-JAX has lower cross-case mean MAE than PyWRF for all three variables. Using WRF with an $8\times16$ decomposition as the reference, the means for $U$, $\theta'$, and $q_r$ at 3\,h are 0.01828\,m\,s$^{-1}$, 0.008371\,K, and $1.702\times10^{-7}$\,kg\,kg$^{-1}$, respectively; the corresponding values in the published PyWRF figure are 0.162\,m\,s$^{-1}$, 0.180\,K, and $2.31\times10^{-6}$\,kg\,kg$^{-1}$. At 6\,h, the corresponding WRF-JAX means are 0.06419\,m\,s$^{-1}$, 0.02697\,K, and $7.844\times10^{-7}$\,kg\,kg$^{-1}$, whereas those for PyWRF are 0.259\,m\,s$^{-1}$, 0.251\,K, and $3.51\times10^{-6}$\,kg\,kg$^{-1}$; the former are $1/4.0$, $1/9.3$, and $1/4.5$ of the latter, respectively.

Figure~\ref{fig:pywrf} places the two WRF difference baselines alongside the error distributions of WRF-JAX and PyWRF, providing a common yardstick for the error comparison.

The evaluations in this study and PyWRF use the same variables, forecast lead times, and statistical definitions. The main differences are that PyWRF has 41 levels and implements only the dynamical core and microphysics parameterization, whereas this study uses 44 levels, all parameterization schemes in the CONUS suite, and the complete dynamical core.

\subsubsection{Evaluation of common diagnostics}
\begin{figure}[ht]
\centering
\includegraphics[width=\textwidth]{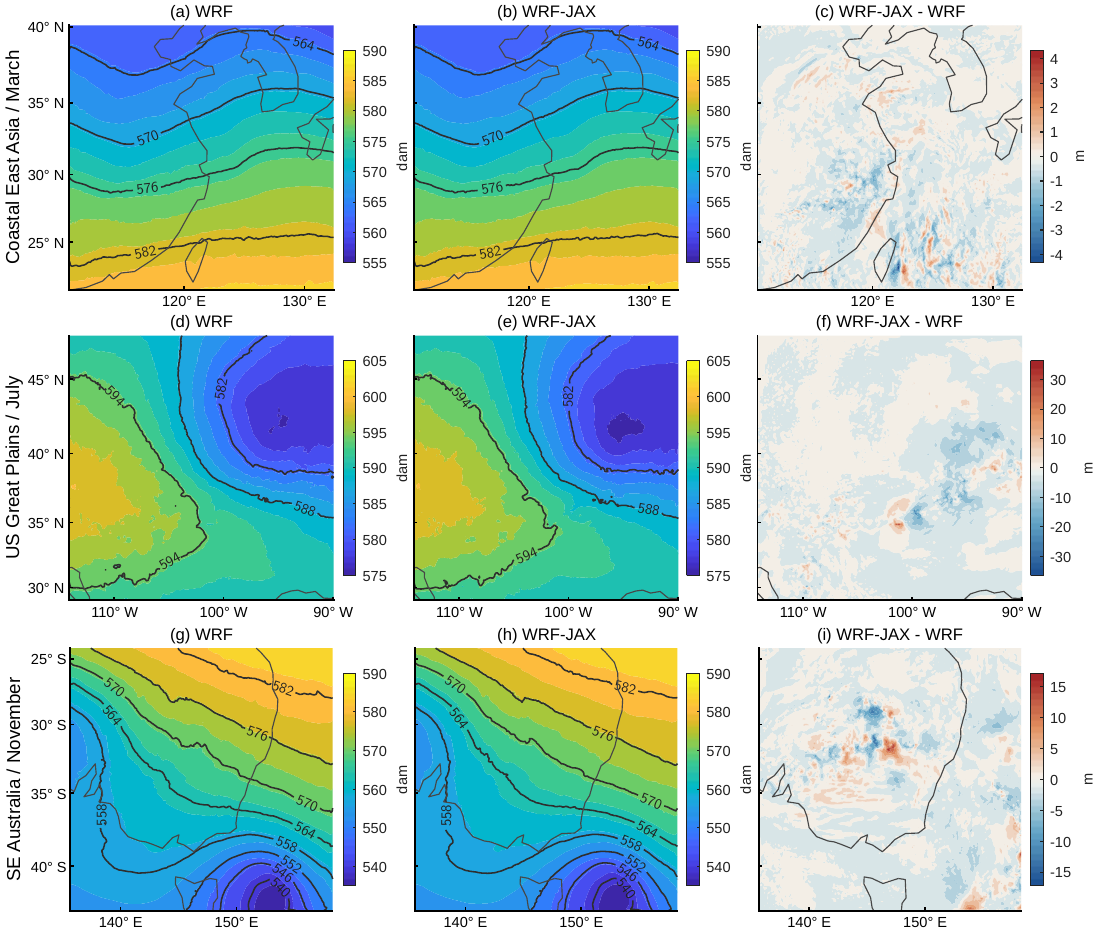}
\caption{500\,hPa geopotential height at 240\,h: East Asian coast in March (a--c), US Great Plains in July (d--f), and southeastern Australia in November (g--i). The three columns show Fortran WRF, WRF-JAX, and WRF-JAX minus WRF, respectively. Model fields are expressed in dam, with a contour interval of 6\,dam; differences are expressed in m. Model fields within each row share a colour scale, and each difference scale is symmetric about zero. Blank areas indicate the absence of a valid pressure-level interpolation interval.}
\label{fig:z500}
\end{figure}

\begin{figure}[p]
\centering
\includegraphics[width=\textwidth]{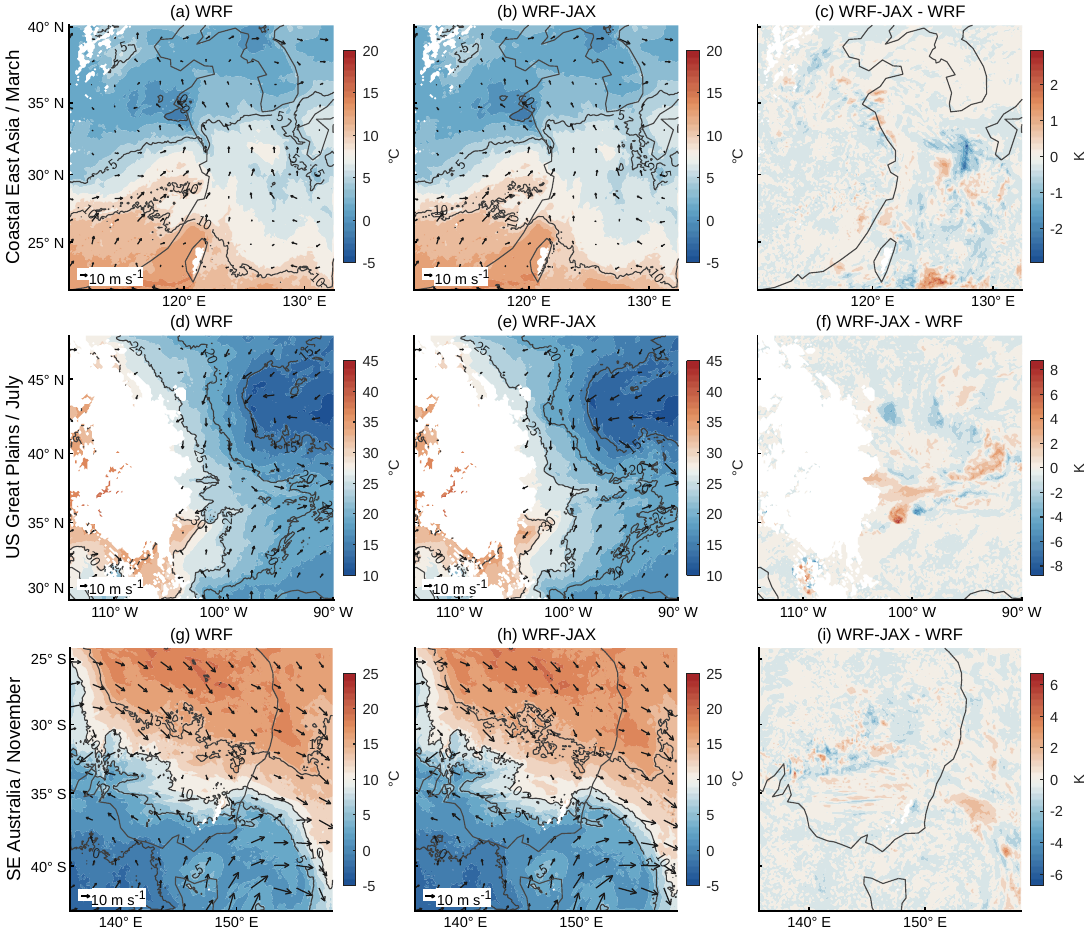}
\caption{850\,hPa air temperature and horizontal wind at 240\,h: East Asian coast in March (a--c), US Great Plains in July (d--f), and southeastern Australia in November (g--i). The three columns show Fortran WRF, WRF-JAX, and WRF-JAX minus WRF, respectively. Temperature in the model fields is expressed in $^{\circ}$C, with a contour interval of 5\,K, and wind vectors have been rotated into geographic coordinates. Difference maps show temperature differences in K, numerically equal to differences in degrees Celsius. WRF and WRF-JAX within each row share the same wind-vector scale, and pressure surfaces below the terrain are masked.}
\label{fig:t850}
\end{figure}

Figures~\ref{fig:z500} and \ref{fig:t850} show the 240\,h fields for the East Asian coast in March, the US Great Plains in July, and southeastern Australia in November. This selection spans three regions and initialization months and includes a convective case with larger wind-field differences. The two model fields in each row use the same colour scale, while difference maps use a symmetric scale centred on zero; geopotential-height contours and low-level wind vectors are used to examine weather-system structure.

Both implementations retain the large-scale 500\,hPa height gradients and the main low-level temperature and wind structures in these cases. At 240\,h, the full-grid height RMSEs for the East Asian coast, US Great Plains, and southeastern Australia are 0.406, 2.532, and 1.683\,m, respectively, with corresponding 850\,hPa temperature RMSEs of 0.284, 0.836, and 0.524\,K. Low-level wind differences are more pronounced in the US Great Plains July case, where the RMSEs of $U_{850}$ and $V_{850}$ reach 2.550 and 2.455\,m\,s$^{-1}$; the corresponding values are 1.000 and 0.899\,m\,s$^{-1}$ for the East Asian coast, and 1.735 and 1.959\,m\,s$^{-1}$ for southeastern Australia. Table~\ref{tab:pressure} presents the pressure-level RMSE and MAE for the three cases at 24 and 240\,h side by side.

Difference maps also reveal local departures that are difficult to discern in the model fields. For example, in the US Great Plains July case, the maximum absolute difference in 500\,hPa height is 50.41\,m, while the domain-wide RMSE is 2.53\,m.

Figures S1--S6 in the Supplement show 500\,hPa geopotential height and 850\,hPa temperature and wind at 24\,h and day ten for all 27 cases, complementing the three 240\,h examples in the main text with a complete cross-region, cross-season comparison. Figures S7--S27 show spatial comparisons of precipitation, radiation, and surface temperature across the nine regions and three initialization months, using common colour scales for the two lead times.

Figures S28--S33 in the Supplement focus on 240\,h, placing the comparison between WRF-JAX and WRF with an $8\times16$ decomposition on the left and the comparison between WRF runs with $8\times16$ and $16\times8$ decompositions on the right for all 27 cases. The two sides use the common valid area and identical model-field and difference colour scales to directly compare the spatial structures of differences at the same time.

\subsubsection{Typhoon track evaluation}
\label{sec:cyclone-tracks}
\begin{figure}[htp]
\centering
\includegraphics[width=\textwidth]{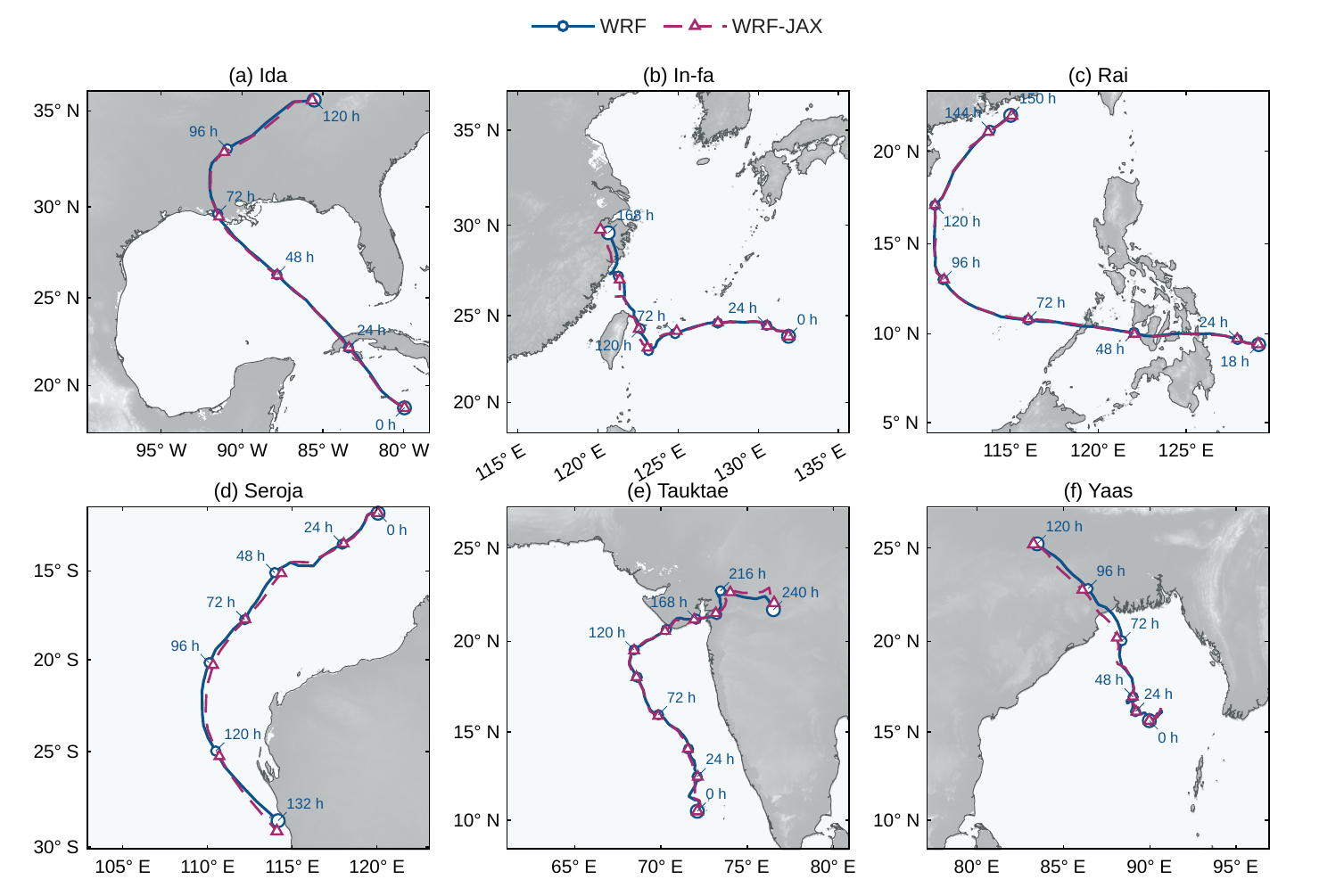}
\caption{Simulated centre tracks for six typhoon cases: (a) Ida, initialized on 27 August 2021; (b) In-fa, initialized on 19 July; (c) Rai, initialized on 15 December; (d) Seroja, initialized on 6 April; (e) Tauktae, initialized on 14 May; (f) Yaas, initialized on 23 May. All initialization times are 00:00 UTC. Blue solid lines denote WRF, and magenta dashed lines denote WRF-JAX. Curves connect centres identified every 3\,h by the GFDL Vortex Tracker. Circles and triangles mark, for the two models respectively, the first point, positions at 24\,h intervals, and the final time shown; the 0\,h markers are obtained from each model's initial fields. Both In-fa tracks are shown through 168\,h, and both Yaas tracks through 120\,h. The two tracks share time labels, with numbers indicating model hours since initialization.}
\label{fig:cyclone-tracks}
\end{figure}

Building on the comparisons of common diagnostics, six typhoon and hurricane cases from around the world in 2021---Ida, In-fa, Rai, Seroja, Tauktae, and Yaas---were further selected to evaluate the ability to simulate real weather events by comparing the tracks simulated by the two implementations. All six cases were integrated for ten days with a 3\,km grid spacing, a $700\times700$ horizontal grid, an 18\,s time step, and the CONUS physics configuration, with output every 3\,h. Figure~\ref{fig:cyclone-tracks} overlays the two simulated tracks for each case.

Cyclone centres are identified using NOAA's GFDL Vortex Tracker~\citep{marchok2021,gfdltracker2026}, with official source-code revision \texttt{6ceb73c}. The tool combines sea-level pressure, geopotential heights at 850 and 700\,hPa, and relative vorticity and wind circulation at these two pressure levels and near the surface to locate cyclone centres. Winds from both models are destaggered and rotated into geographic coordinates, pressure-level fields are interpolated in log-pressure coordinates, and the required fields are mapped to a regular $0.025^{\circ}$ latitude--longitude grid using the same bilinear interpolation. Each case is initialized with the earliest operational centre position within the common valid area in the TC Vitals archive supplied with the tool, after which tracking proceeds separately using each model's fields.

The Ida and Rai tracks produced by the two implementations are very close; the main directions of motion and track shapes of In-fa, Seroja, Tauktae, and Yaas are also consistent, with differences appearing as shifts in centre positions and turning details during some periods. This comparison extends the instantaneous meteorological-variable comparisons in Figs.~\ref{fig:z500} and \ref{fig:t850} to the continuous evolution of weather systems.

\subsection{Computational performance}
\label{sec:performance}
\subsubsection{Computational throughput of ten-day integrations}
The CPU used for the Fortran timing tests is an Intel Xeon CPU Max 9462. WRF is compiled with gfortran 13.3.0 and ifort 2021.10.0, respectively.

The mean sustained integration times of 128-process Fortran WRF compiled with gfortran and ifort are 205.538 and 183.963\,s per model hour, respectively, compared with 174.951\,s per model hour for WRF-JAX, corresponding to speedups of 1.175 and 1.052 (Table~\ref{tab:performance}). A single A800 thus achieves sustained throughput comparable to that of the measured 128-process CPU configuration.

\subsubsection{Sustained integration speed with different CPU configurations}
\label{sec:singlecore}
The CPU used for the timing tests is an Intel Xeon CPU Max 9462. WRF 4.5 is built in \texttt{dmpar} mode. The gfortran toolchain comprises gfortran 13.3.0 and Open MPI 5.0.8, with the main Fortran optimization options \texttt{-O2 -ftree-vectorize -funroll-loops}; the ifort toolchain comprises ifort 2021.10.0 and Intel MPI 2021.10, using the default \texttt{-O3} and retaining \texttt{-fp-model precise}. All these options are the default WRF compilation settings.

The initial and boundary fields are taken from a June 2021 case centred at $28^\circ\,\mathrm{N}$, $75^\circ\,\mathrm{W}$, with a $700\times700$ horizontal grid, 3\,km grid spacing, 44 mass levels, and the complete CONUS physics suite enabled.

The time step remains 18\,s, and the radiation interval is set to 10\,min. Each run covers 86400\,s of model time, with disk-write time subtracted. WRF process CPU time is sampled every 10\,s, and mean utilization during integration is calculated from the ratio of CPU-time increments to wall-clock-time increments.

On the Xeon CPU Max 9462 platform, the mean sustained integration times of the single-process gfortran and ifort builds are 13\,464.221 and 11\,080.220\,s per model hour, respectively, with the ifort build running 1.215 times as fast as the gfortran build.

To extend the CPU-platform performance evaluation, single-process and 64-process timing tests of the gfortran and ifort builds were also conducted on an AMD EPYC 7542 processor. Both WRF and WPS are version 4.5, and the grid size, resolution, vertical levels, physics suite, time step, and radiation interval are the same as in the tests above. The gfortran 13.3.0 and ifort 2021.10.0 builds retain their default \texttt{-O2} and \texttt{-O3} optimization configurations, respectively.

Sustained integration time is summarized using means for both CPU platforms (Table~\ref{tab:performance}). Mean CPU utilization normalized by the MPI process count is close to 100\,\% in all cases. The table shows that ifort builds run faster than gfortran builds and that, because of performance differences among CPU hardware, speed increases more nearly linearly with core count on some CPUs, whereas others reach the MPI communication bottleneck sooner. These findings are consistent with known behaviour.

\begin{table}[ht]
\caption{Summary of sustained integration performance. The results for a single A800 and the 128-process gfortran run on Xeon CPU Max 9462 are means over the same set of 27 paired cases in the ten-day experiments, timed over 3--240\,h. All timings exclude output. Parallelism in CPU rows denotes the number of MPI processes, and CPU utilization is normalized by this number; 100\,\% means that each process occupies one logical CPU on average. CPU utilization is included to verify exclusive CPU use during timing tests and rule out interference from other processes in the timing results. GPU utilization is the mean of samples collected during sustained integration windows without data input/output, excluding initialization and compilation, and averaged with equal weight across the 27 cases. Dashes denote inapplicable entries.}
\label{tab:performance}
\centering
\small
\setlength{\tabcolsep}{2.5pt}
\begin{tabularx}{\textwidth}{@{}>{\raggedright\arraybackslash}p{1.55cm}>{\raggedright\arraybackslash}p{1.7cm}>{\raggedright\arraybackslash}p{1.4cm}>{\centering\arraybackslash}p{1.6cm}*{3}{>{\raggedleft\arraybackslash}X}*{2}{>{\raggedleft\arraybackslash}p{1.6cm}}>{\raggedleft\arraybackslash}p{1.85cm}@{}}
\tophline
\multirow{3}{*}[-1mm]{Model} & \multirow{3}{1.7cm}[-1mm]{Computing platform} & \multirow{3}{*}[-1mm]{Compiler} & \multirow{3}{1.6cm}[-1mm]{\centering Parallelism} & \multicolumn{3}{c}{\shortstack{Sustained integration time\\(s per model hour)}} & \multicolumn{2}{c}{\shortstack{Mean CPU\\utilization (\%)}} & \multirow{3}{1.85cm}[-1mm]{\raggedleft Mean GPU\newline utilization (\%)} \\
\cline{5-7}\cline{8-9}
 & & & & Single precision & Double precision & Mixed precision & Single precision & Double precision & \\
\middlehline
WRF-JAX & A800 & jax & Single GPU & - & - & 174.951 & -- & - & 96.50 \\
\middlehline
\multirow{6}{1.55cm}{WRF} & \multirow{6}{1.7cm}{\raggedright Intel Xeon CPU Max 9462} & gfortran & 1 & 13464.221 & 27\,754.993 & - & 99.99 & 99.99 & -- \\
 & & ifort & 1 & 11080.220 & 17\,609.155 & - & 99.99 & 99.98 & -- \\
 & & gfortran & 64 & 377.611 & 974.685 & - & 99.57 & 99.64 & -- \\
 & & ifort & 64 & 334.624 & 884.393 & - & 99.58 & 99.62 & -- \\
 & & gfortran & 128 & 205.538 & 631.898 & - & 99.43 & 99.57 & -- \\
 & & ifort & 128 & 183.963 & 302.606 & - & 99.26 & 99.58 & -- \\
\middlehline
\multirow{4}{1.55cm}{WRF} & \multirow{4}{1.7cm}{\raggedright AMD EPYC 7542} & gfortran & 1 & 15\,389.550 & 25\,465.794 & - & 99.98 & 99.98 & -- \\
 & & ifort & 1 & 12\,692.801 & 18\,219.634 & - & 99.99 & 99.99 & -- \\
 & & gfortran & 64 & 1452.563 & 2942.014 & - & 99.99 & 99.94 & -- \\
 & & ifort & 64 & 1231.181 & 2618.264 & - & 99.99 & 99.98 & -- \\
\bottomhline
\end{tabularx}
\end{table}

\subsubsection{Performance comparison with existing work}
Table~3 of \citet{chu2026} reports six-hour PyWRF integration times on an A100 of 46\,min in single precision and 73\,min in double precision, corresponding to 460 and 730\,s per model hour, respectively. Its reported speedups of 12.4 and 7.8 both use double-precision Fortran WRF on a single Hygon 7390 CPU core as the baseline, for which the six-hour integration takes 572\,min; the 12.4-fold speedup compares single precision on the GPU with double precision on the CPU. The public PyWRF repository also states that ``computational performance is far below the Fortran reference''~\citep{pywrf2026}.

The Fortran reference baselines in this study run in single precision on Intel Xeon CPU Max 9462, using gfortran with \texttt{-O2} and ifort with \texttt{-O3} optimization, respectively, and 128 MPI processes in parallel. With the complete CONUS physics configuration, a single A800 achieves sustained integration speedups of 1.175 and 1.052 relative to these two compiler configurations (Table~\ref{tab:performance}). This demonstrates that, while supporting automatic differentiation through the fully coupled dynamics and physics, WRF-JAX already achieves the sustained forward-integration throughput of compiler-optimized, 128-process Fortran WRF on a single GPU.

The public documentation of \texttt{wrf\_gpu} reports a speedup of approximately 1.53 relative to 12 CPU MPI processes for a particular version and case, together with scaling results on other grids~\citep{wrfgpu2026}. ArWen and its derivative WOOF use a mixed Python, CuPy, and CUDA technology stack~\citep{arwen2026,woof2026}. These projects differ in hardware, physics coverage, integration duration, and timing definitions; the performance conclusions in this study are based on paired measurements of WRF-JAX and Fortran WRF.

\subsection{Validation of WRF-JAX differentiability}
\label{sec:ad-results}
Under full coupling of dynamics and physics, WRF-JAX backpropagates gradients through 100 steps to the initial UVTQ fields, obtaining finite gradient components for the entire three-dimensional fields of all four variables in a single calculation. The baseline objective value is 33.009319\,K. This reverse-mode window contains four radiation updates and continuous microphysics, land-surface, surface-layer, boundary-layer, and convective processes, incorporating their responses across time steps into the initial-field gradients.

Table~\ref{tab:reverse100-summary} presents finite-difference comparisons for 500 full-field directions per variable. Across the three amplitudes, the median relative differences are 2.58--5.22\,\% for $U$, 2.94--6.49\,\% for $V$, 7.86--10.06\,\% for $T$, and 3.51--3.91\,\% for $Q$. The reverse-mode projections and finite differences have matching signs in all 6000 amplitude checks. For $Q$, the differences are below 5\,\% for all 500 directions at each of the three amplitudes; for $U$ and $V$, the mean relative differences at $h=0.3$\,m\,s$^{-1}$ are 4.16\,\% and 4.39\,\%, respectively.

\begin{table}[ht]
\caption{Full-field directional finite-difference evaluation of 100-step backpropagation. The table reports the 25th percentile, median, 75th percentile, and arithmetic mean of the relative differences in directional derivatives, $100r(h)$. $T$ denotes the initial dry potential temperature perturbation, and $Q$ uses non-negative directions and one-sided differences.}
\label{tab:reverse100-summary}
\centering
\begin{tabularx}{\textwidth}{llrr*{4}{>{\raggedleft\arraybackslash}X}}
\tophline
Initial variable & Amplitude units & Amplitude $h$ & Directions & 25th percentile\newline (\%) & Median\newline (\%) & 75th percentile\newline (\%) & Mean\newline (\%) \\
\middlehline
$U$ & m\,s$^{-1}$ & 0.3 & 500 & 1.10 & 2.58 & 5.50 & 4.16 \\
 &  & 0.1 & 500 & 1.58 & 3.40 & 6.75 & 5.01 \\
 &  & 0.03 & 500 & 2.02 & 5.22 & 10.22 & 6.87 \\
\middlehline
$V$ & m\,s$^{-1}$ & 0.3 & 500 & 1.17 & 2.94 & 5.47 & 4.39 \\
 &  & 0.1 & 500 & 1.89 & 4.20 & 8.45 & 6.03 \\
 &  & 0.03 & 500 & 2.72 & 6.49 & 12.48 & 8.43 \\
\middlehline
$T$ & K & 0.1 & 500 & 4.55 & 10.06 & 19.98 & 15.41 \\
 &  & 0.03 & 500 & 3.57 & 7.86 & 15.47 & 12.30 \\
 &  & 0.01 & 500 & 4.10 & 8.96 & 18.79 & 13.63 \\
\middlehline
$Q$ & kg\,kg$^{-1}$ & $1\times10^{-5}$ & 500 & 3.35 & 3.51 & 3.71 & 3.52 \\
 &  & $3\times10^{-6}$ & 500 & 3.74 & 3.91 & 4.04 & 3.88 \\
 &  & $1\times10^{-6}$ & 500 & 3.26 & 3.53 & 3.81 & 3.53 \\
\bottomhline
\end{tabularx}
\end{table}

This implementation organizes the WRF-JAX forecast as a computational process that supports backpropagation: a single reverse calculation simultaneously provides gradients for all initial control components, while finite differences sample these directional responses for evaluation. The resulting approximate gradients can serve as update signals for initial-field inversion and trainable correction modules; through the chain rule for initial-field or parameter mappings, the final-state objective can be connected to upstream trainable variables, providing the differentiation interface needed for end-to-end optimization.

\section{Discussion}
\subsection{Main contributions of WRF-JAX}
The main advance of this work is the combination, within WRF-JAX, of a coupled physics configuration derived from WRF, GPU-accelerated execution, and validation across multiple periods and regions. Radiation, land surface, surface processes, turbulence, and convection are all enabled in the accuracy and speed tests, distinguishing this study from demonstrations limited to advection or microphysics. Explicit states and separate execution scheduling also make component exchanges easier to inspect.

WRF-JAX was developed through a progressive porting workflow. Coverage analysis focuses on the selected dynamics and physics configuration, call-dependency ordering supports prior verification of low-level components, and temporary Fortran bridges maintain full coupling during module-by-module replacement. Difference checks with identical inputs separately locate issues in formulas, interfaces, and execution organization, followed by complete integrations to examine feedback. This workflow embeds numerical checks throughout component porting and performance optimization in WRF-JAX.

The ten-day experiments continuously test time-varying boundary updates and the long-term evolution of land-surface and physics states. The 27 cases span different regions and seasons, providing richer numerical evidence than short integrations of a single case. Native-grid errors, evaluations of common diagnostics, and the original model's error baseline complement one another to provide error assessment from multiple perspectives.

\subsection{Numerical consistency}
None of the existing efforts has established a WRF internal-difference baseline, and their errors are in fact many times larger than this baseline. PyWRF is used as the cited comparison case in this study; \texttt{wrf\_gpu} and ArWen are not included in the evaluation for reasons including evaluation settings that differ from those used here, errors larger than those of PyWRF, and the absence of publicly available detailed evaluation results. The WRF internal numerical-difference baseline in this study provides a criterion for assessing model numerical consistency. Although wind, temperature, and moisture RMSEs increase with lead time, WRF-JAX remains close to the original model's difference baseline throughout, with median case-wise ratios of 1.02--1.04 on day ten. The baselines constructed by the two methods are essentially the same. After comparing multiple baselines, the ten-day evaluation uses a fixed $8\times16$ and $16\times8$ decomposition pair because of computational cost, while the six-hour short-term forecast validation uses both the parallel-decomposition difference baseline and the ensemble-difference baseline from floating-point perturbations to the initial fields.

Although WRF-JAX remains close to the original model's difference baseline throughout and substantially outperforms several existing efforts, its errors are slightly above the baseline. Such differences are nevertheless tolerable in forecasting real weather events, and the model does have practical forecasting capability; the typhoon track evaluation further demonstrates consistency between WRF-JAX and the original WRF model. Differences in programming language, framework, and execution hardware make bitwise agreement between WRF-JAX and the original WRF model difficult to achieve. However, the model itself has inherent biases, and the WRF-JAX implementation need not reproduce the original model's results bit for bit, as doing so would inherit those biases. Future work using a forecast-capable model such as WRF-JAX as a foundation for end-to-end optimization of model parameters, or for end-to-end optimization after replacing physical parameterizations with AI models, can further improve forecasts of real weather events. Forecasting capability can then be evaluated against actual observations rather than against a WRF model with inherent biases.

\subsection{Future work: smooth AI parameterizations for end-to-end training}
Section~\ref{sec:ad-results} demonstrates 100-step backpropagation under full physics coupling and quantifies the quality of its approximate gradients using full-field Gaussian random directions. In the evaluation of 500 directions per variable, the median relative differences across variables and amplitudes range from 2.58 to 10.06\,\%. Backpropagation returns gradients to all initial control vectors, enabling the forecast process to be connected to gradient-based initial-field adjustment and end-to-end optimization.

Existing physical parameterizations contain threshold tests, discrete table lookups, and process switching. Backpropagation differentiates along the currently active branch, whereas finite-difference results depend on the perturbation amplitude $h$. When $h$ is large, a perturbation may cross branch boundaries or trigger different physical processes, so the resulting finite difference includes process switching and nonlinear responses and thus deviates from the local derivative at the current state. When $h$ is too small, the objective change caused by the perturbation may be overwhelmed by floating-point round-off errors in single-precision integration, and subtraction in the finite difference and division by $h$ further amplify their effects. Consequently, for some states and perturbation directions, there may be no suitable $h$ that both avoids crossing branch boundaries and keeps the finite-difference signal sufficiently above round-off error, making a stable, reliable finite-difference reference difficult to obtain.

In addition, the approximate derivatives of continuous physical responses used for discrete radiation table lookups also affect agreement between reverse-mode gradients and finite differences. Differences in gradient checks are therefore influenced jointly by approximate differentiation rules, model non-smoothness, and finite-difference numerical errors; they cannot all be attributed to errors in backpropagation itself.

Approximate gradients with some discrepancies can still be used for optimization, where their role is to provide effective update directions for the objective function. All 6000 directional derivatives checked in this study agree in sign, indicating that, along these directions, the reverse-mode gradients capture whether finite perturbations increase or decrease the objective. This supports their use as update signals for initial-field adjustment and trainable modules.

Future work will progressively replace the existing physical parameterizations with smooth, differentiable AI modules to construct a regional model coupling the ARW dynamical core with an entirely AI-based parameterization suite. Reducing threshold tests, discrete table lookups, and process switching in the corresponding physical processes, and directly differentiating the continuous calculations that replace them, is expected to substantially reduce the gradient discrepancies caused by these non-smooth and discrete treatments. This should yield smoother and more accurate gradients across time steps while widening the range of usable perturbation amplitudes in finite-difference checks. The degree of improvement will be tested using finite differences, and the practical effectiveness of the gradients for optimization will be evaluated through end-to-end training.

\conclusions
WRF-JAX v1.1.28 implements the WRF 4.5 dynamics and the evaluated CONUS coupled physics suite in JAX. Starting with coverage and call-dependency analysis, development progressively translated the dynamics and physics components through temporary Fortran bridges, module-by-module replacement, and difference checks with identical inputs. Staged compilation, bounded column batches, and streaming output support large-domain integrations on a single GPU.

The 27 ten-day experiments provide a numerical evaluation across regions and seasons. The WRF internal numerical-difference baseline established for all cases shows that the wind, temperature, and moisture errors in WRF-JAX are of similar magnitude to the reference model's own long-term numerical differences: the median case-wise RMSE ratios are 1.05--1.07 at 24\,h and 1.02--1.04 at 240\,h. On this basis, box-plot comparisons with published PyWRF results, using the same variables and cross-case statistics at individual lead times, show that the experiment set in this study has lower mean MAEs for zonal wind, perturbation potential temperature, and rainwater at 1--6\,h.

The mean sustained integration time on a single A800 is 175.0\,s per model hour, corresponding to speedups of 1.17 and 1.05 relative to 128-process single-precision WRF runs compiled with gfortran and ifort, respectively. WRF-JAX thus achieves practical computational throughput for a fully coupled regional weather model on a single GPU and, through explicit states and reproducible evaluation, provides a foundation for further model development and integration into the Python ecosystem.

In the 100-step backpropagation evaluation of WRF-JAX, all gradient components of the complete three-dimensional fields of the four initial UVTQ variables can be obtained in a single reverse calculation. The evaluation of 500 full-field directions per variable at three amplitudes provides 6000 finite-difference comparisons: median relative differences across variables and amplitudes range from 2.58 to 10.06\,\%, and all derivative signs agree. These results combine coupled regional forecasting with approximate gradient computation, providing a scientific computing foundation for applications such as solving CNOP problems and end-to-end training.

\clearpage
\appendix
\section{Additional numerical statistics}
Tables~\ref{tab:rmse} to \ref{tab:nrmse} report cross-case medians at specified lead times. Mixing-ratio units are converted for readability, and each absolute-error statistic uses all 27 cases. For both comparisons, normalized shortwave errors at 24 and 240\,h use 14 and 13 cases, respectively, because the root-mean-square amplitude of the reference field is zero in the remaining cases; all other normalized errors in the tables use all 27 cases.

\begin{table}[htp]
\caption{Cross-case median RMSEs at 24 and 240\,h. WRF-JAX vs WRF and WRF internal differences are calculated as the RMSEs of WRF-JAX and WRF with a $16\times8$ decomposition, respectively, relative to the same WRF reference with an $8\times16$ decomposition, each including 27 cases.}
\label{tab:rmse}
\centering
\begin{tabular*}{\textwidth}{@{\extracolsep{\fill}}llrrrr@{}}
\tophline
Variable & Units & \multicolumn{2}{c}{24\,h} & \multicolumn{2}{c}{240\,h} \\
 & & \shortstack{WRF-JAX\\vs WRF} & \shortstack{WRF internal\\differences} & \shortstack{WRF-JAX\\vs WRF} & \shortstack{WRF internal\\differences} \\
\middlehline
$U$ & m s$^{-1}$ & 0.589 & 0.565 & 1.04 & 1.03 \\
$V$ & m s$^{-1}$ & 0.572 & 0.539 & 0.980 & 0.946 \\
$\theta'$ & K & 0.232 & 0.209 & 0.396 & 0.391 \\
$q_v$ & g kg$^{-1}$ & 0.254 & 0.232 & 0.457 & 0.476 \\
$q_r$ & mg kg$^{-1}$ & 13.1 & 13.6 & 45.2 & 45.7 \\
$T_2$ & K & 0.293 & 0.242 & 0.424 & 0.404 \\
$q_2$ & g kg$^{-1}$ & 0.291 & 0.260 & 0.527 & 0.498 \\
$U_{10}$ & m s$^{-1}$ & 0.680 & 0.659 & 0.992 & 0.957 \\
$V_{10}$ & m s$^{-1}$ & 0.703 & 0.677 & 1.02 & 1.00 \\
$p_s$ & hPa & 0.0710 & 0.0613 & 0.140 & 0.139 \\
RAINNC & mm & 1.19 & 1.18 & 13.5 & 12.6 \\
RAINC & mm & 0.368 & 0.330 & 2.24 & 2.23 \\
SWDNB & W m$^{-2}$ & 0.00136 & 0.00168 & 0 & 0 \\
LWDNB & W m$^{-2}$ & 10.3 & 9.23 & 13.4 & 13.2 \\
\bottomhline
\end{tabular*}
\end{table}

\begin{table}[htp]
\caption{Cross-case median MAEs, using the same two comparisons as in Table~\ref{tab:rmse}.}
\label{tab:mae}
\centering
\begin{tabular*}{\textwidth}{@{\extracolsep{\fill}}llrrrr@{}}
\tophline
Variable & Units & \multicolumn{2}{c}{24\,h} & \multicolumn{2}{c}{240\,h} \\
 & & \shortstack{WRF-JAX\\vs WRF} & \shortstack{WRF internal\\differences} & \shortstack{WRF-JAX\\vs WRF} & \shortstack{WRF internal\\differences} \\
\middlehline
$U$ & m s$^{-1}$ & 0.291 & 0.286 & 0.619 & 0.581 \\
$V$ & m s$^{-1}$ & 0.291 & 0.272 & 0.604 & 0.590 \\
$\theta'$ & K & 0.126 & 0.115 & 0.249 & 0.250 \\
$q_v$ & g kg$^{-1}$ & 0.104 & 0.0901 & 0.246 & 0.235 \\
$q_r$ & mg kg$^{-1}$ & 0.779 & 0.748 & 1.92 & 1.97 \\
$T_2$ & K & 0.143 & 0.125 & 0.254 & 0.246 \\
$q_2$ & g kg$^{-1}$ & 0.169 & 0.143 & 0.350 & 0.333 \\
$U_{10}$ & m s$^{-1}$ & 0.412 & 0.357 & 0.604 & 0.590 \\
$V_{10}$ & m s$^{-1}$ & 0.407 & 0.356 & 0.632 & 0.615 \\
$p_s$ & hPa & 0.0395 & 0.0342 & 0.0799 & 0.0759 \\
RAINNC & mm & 0.174 & 0.170 & 5.53 & 5.27 \\
RAINC & mm & 0.0862 & 0.0782 & 1.03 & 0.902 \\
SWDNB & W m$^{-2}$ & $1.96\times10^{-5}$ & $2.20\times10^{-5}$ & 0 & 0 \\
LWDNB & W m$^{-2}$ & 3.43 & 3.14 & 5.99 & 5.81 \\
\bottomhline
\end{tabular*}
\end{table}

\begin{table}[htp]
\caption{Cross-case median nRMSEs, using the same two comparisons as in Table~\ref{tab:rmse}, both normalized by the root-mean-square amplitude of the same reference field from the $8\times16$ decomposition according to Eq.~(\ref{eq:nrmse}).}
\label{tab:nrmse}
\centering
\begin{tabular*}{\textwidth}{@{\extracolsep{\fill}}llrrrr@{}}
\tophline
Variable & Units & \multicolumn{2}{c}{24\,h} & \multicolumn{2}{c}{240\,h} \\
 & & \shortstack{WRF-JAX\\vs WRF} & \shortstack{WRF internal\\differences} & \shortstack{WRF-JAX\\vs WRF} & \shortstack{WRF internal\\differences} \\
\middlehline
$U$ & \% & 5.39 & 5.15 & 11.8 & 11.2 \\
$V$ & \% & 10.8 & 10.6 & 19.7 & 17.5 \\
$\theta'$ & \% & 0.426 & 0.384 & 0.779 & 0.803 \\
$q_v$ & \% & 3.74 & 3.44 & 7.02 & 7.06 \\
$q_r$ & \% & 107 & 105 & 120 & 120 \\
$T_2$ & \% & 0.0974 & 0.0807 & 0.144 & 0.136 \\
$q_2$ & \% & 2.29 & 2.21 & 3.98 & 3.55 \\
$U_{10}$ & \% & 20.0 & 17.6 & 24.8 & 25.2 \\
$V_{10}$ & \% & 18.9 & 16.2 & 34.5 & 33.5 \\
$p_s$ & \% & 0.00740 & 0.00610 & 0.0154 & 0.0154 \\
RAINNC & \% & 54.2 & 49.0 & 50.6 & 46.4 \\
RAINC & \% & 37.8 & 35.1 & 39.9 & 36.0 \\
SWDNB & \% & 17.7 & 17.1 & 20.4 & 20.1 \\
LWDNB & \% & 2.92 & 2.79 & 3.79 & 3.75 \\
\bottomhline
\end{tabular*}
\end{table}

\begin{table}[htp]
\caption{Pressure-level differences at 24 and 240\,h for the three cases in Figs.~\ref{fig:z500} and \ref{fig:t850}. Statistics use all horizontal columns with valid bracketing interpolation levels in both simulations. All 488\,601 columns are used at 500\,hPa; at 850\,hPa, the numbers of common valid columns for the East Asian coast, US Great Plains, and southeastern Australia are 483\,911, 360\,229, and 488\,483 at 24\,h, and 483\,839, 362\,462, and 488\,161 at 240\,h, respectively.}
\label{tab:pressure}
\centering
\begin{tabular*}{\textwidth}{@{\extracolsep{\fill}}lllrrrr@{}}
\tophline
Case & Diagnostic & Units & \multicolumn{2}{c}{24 h} & \multicolumn{2}{c}{240 h} \\
 & & & RMSE & MAE & RMSE & MAE \\
\middlehline
East Asian coast / March & $Z_{500}$ & m & 0.318 & 0.172 & 0.406 & 0.261 \\
 & $T_{850}$ & K & 0.187 & 0.0978 & 0.284 & 0.172 \\
 & $U_{850}$ & m s$^{-1}$ & 0.520 & 0.253 & 1.00 & 0.581 \\
 & $V_{850}$ & m s$^{-1}$ & 0.479 & 0.240 & 0.899 & 0.546 \\
US Great Plains / July & $Z_{500}$ & m & 0.994 & 0.537 & 2.53 & 1.59 \\
 & $T_{850}$ & K & 0.368 & 0.188 & 0.836 & 0.492 \\
 & $U_{850}$ & m s$^{-1}$ & 1.33 & 0.775 & 2.55 & 1.61 \\
 & $V_{850}$ & m s$^{-1}$ & 1.28 & 0.753 & 2.46 & 1.59 \\
Southeastern Australia / November & $Z_{500}$ & m & 0.271 & 0.143 & 1.68 & 1.02 \\
 & $T_{850}$ & K & 0.140 & 0.0788 & 0.524 & 0.336 \\
 & $U_{850}$ & m s$^{-1}$ & 0.582 & 0.302 & 1.74 & 1.06 \\
 & $V_{850}$ & m s$^{-1}$ & 0.558 & 0.297 & 1.96 & 1.13 \\
\bottomhline
\end{tabular*}
\end{table}

\begin{figure}[htp]
\centering
\includegraphics[width=\textwidth]{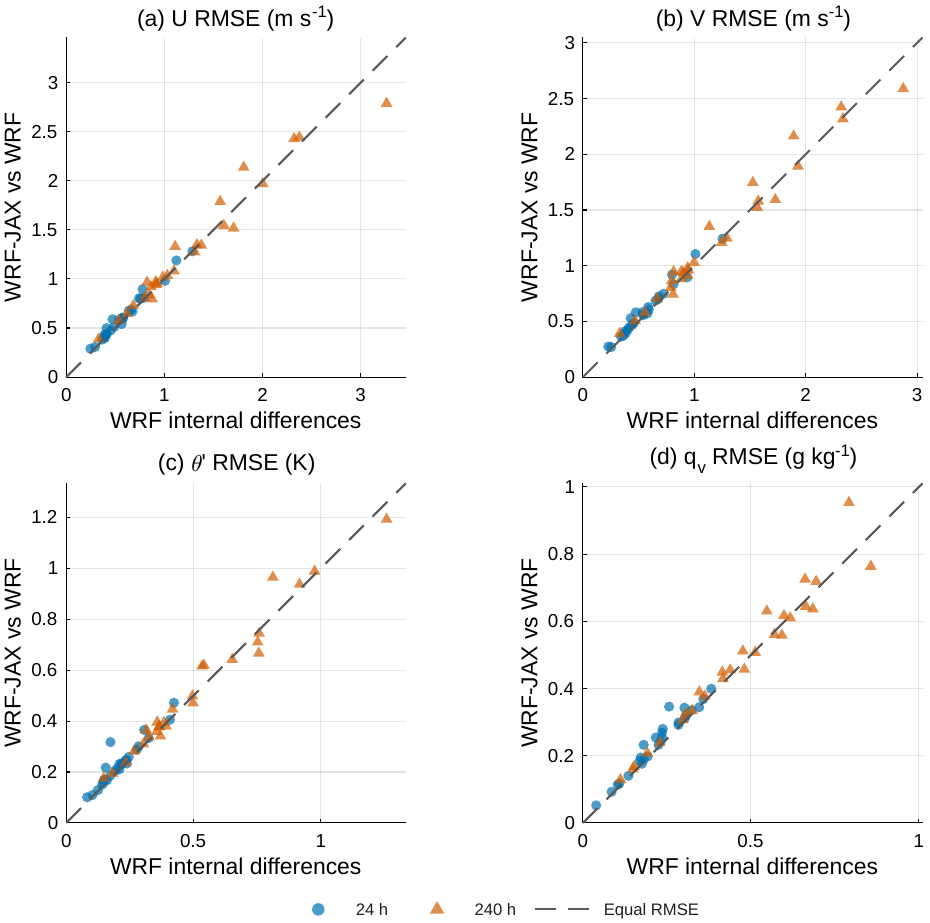}
\caption{Paired case-wise RMSE comparisons for the 27 cases: (a) $U$; (b) $V$; (c) $\theta'$; (d) water vapour. The horizontal axis denotes WRF internal differences and the vertical axis denotes WRF-JAX vs WRF, calculated as the RMSEs of WRF with a $16\times8$ decomposition and WRF-JAX, respectively, relative to the same WRF reference with an $8\times16$ decomposition. Blue circles indicate 24\,h and orange triangles indicate 240\,h, each symbol type containing 27 cases; dashed lines indicate equal RMSEs for the two comparisons. Units are given in the panel titles.}
\label{fig:mpi}
\end{figure}

\begin{figure}[p]
\centering
\includegraphics[width=0.96\textwidth]{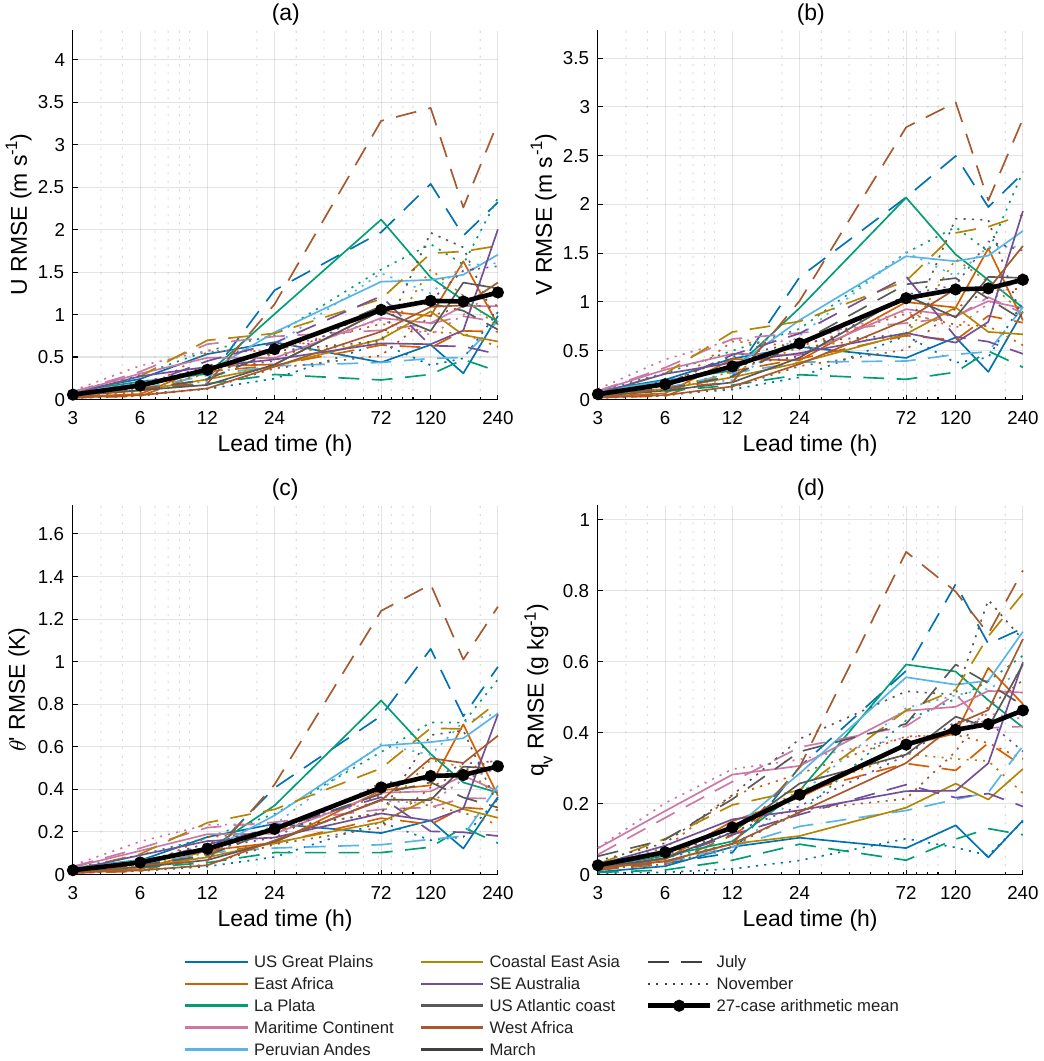}
\caption{Case-wise RMSE evolution of WRF internal differences: (a) $U$; (b) $V$; (c) $\theta'$; (d) water vapour. Differences are calculated as the RMSE of WRF with a $16\times8$ decomposition relative to WRF with an $8\times16$ decomposition. Each panel contains 27 case curves, with colours distinguishing the nine regions and solid, dashed, and dotted lines denoting March, July, and November initializations, respectively. The thick black line with circles is the arithmetic mean of the 27 case RMSEs at each lead time. The horizontal axis is logarithmic; each curve connects the measured metrics at 3, 6, 12, 24, 72, 120, 168, and 240\,h.}
\label{fig:mpi-case-lines}
\end{figure}

\begin{figure}[p]
\centering
\includegraphics[width=\textwidth]{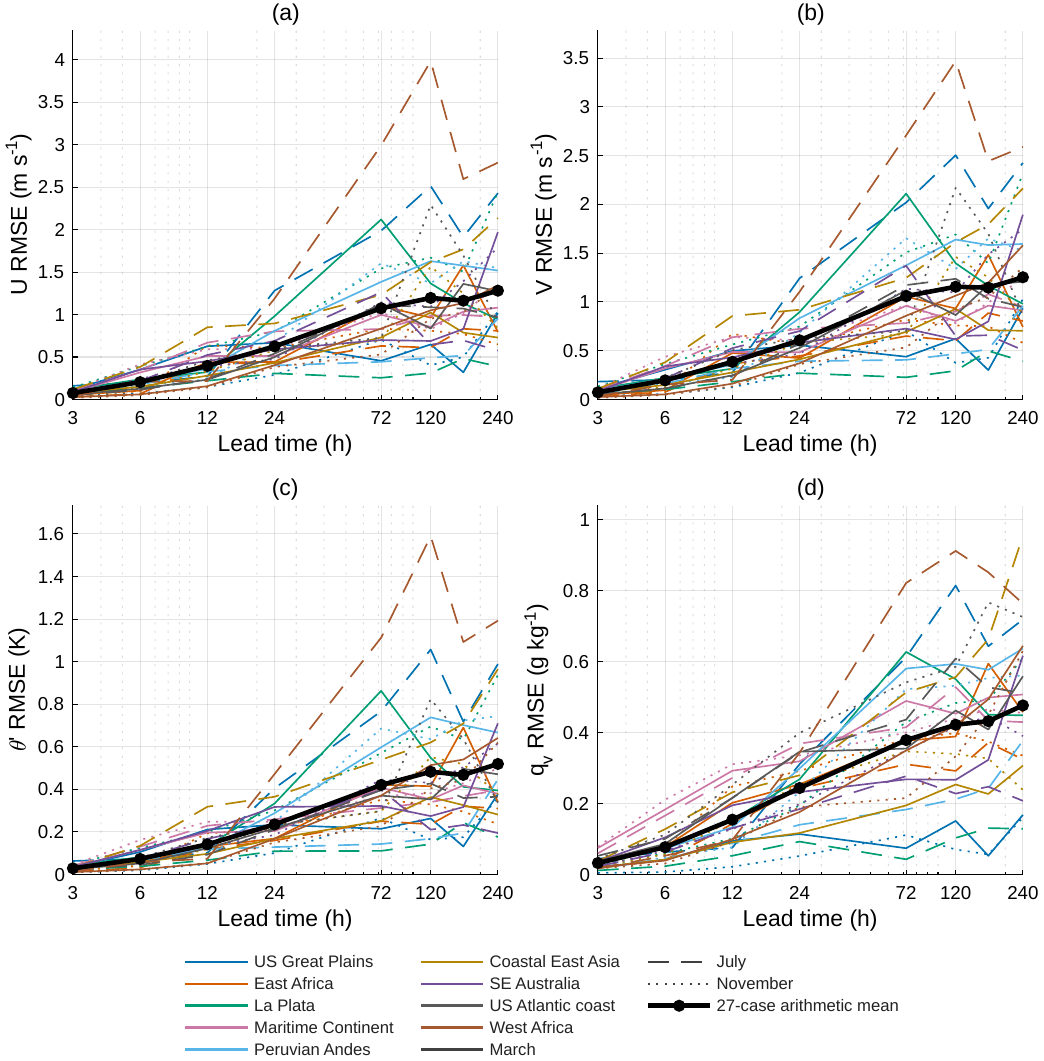}
\caption{Case-wise RMSE evolution of WRF-JAX vs WRF, calculated relative to WRF with an $8\times16$ decomposition. Variables, colours and line styles for the 27 cases, arithmetic-mean markers, and axis ranges are identical to those in Fig.~\ref{fig:mpi-case-lines}.}
\label{fig:jax-case-lines}
\end{figure}

Figures~\ref{fig:mpi-case-lines} and \ref{fig:jax-case-lines} show the case-wise RMSE evolution for the two comparisons, respectively. Each curve connects the eight common analysis times, and the thick black line is the equally weighted arithmetic mean of the RMSEs for the 27 cases. The two figures use matching region colours, month line styles, and vertical-axis ranges to facilitate case-wise comparisons. At 240\,h, the mean RMSEs of $U$, $V$, $\theta'$, and water vapour in the WRF internal numerical-difference baseline are 1.260\,m\,s$^{-1}$, 1.227\,m\,s$^{-1}$, 0.5081\,K, and 0.4625\,g\,kg$^{-1}$, respectively; the corresponding WRF-JAX means are 1.281\,m\,s$^{-1}$, 1.253\,m\,s$^{-1}$, 0.5193\,K, and 0.4764\,g\,kg$^{-1}$.

\begin{figure}[p]
\centering
\includegraphics[width=\textwidth]{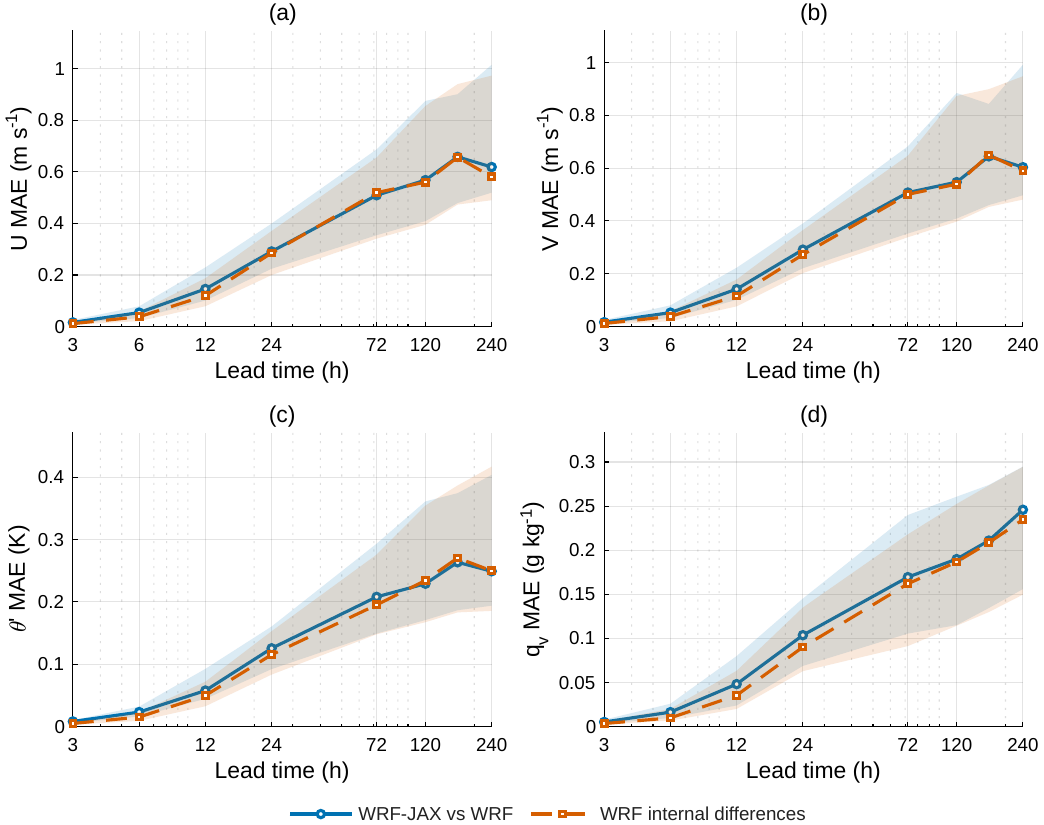}
\caption{Evolution of full-field MAE for the 27 cases: (a) $U$; (b) $V$; (c) $\theta'$; (d) water vapour. Blue lines denote WRF-JAX vs WRF, and orange lines denote WRF internal differences, calculated as the MAEs of WRF-JAX and WRF with a $16\times8$ decomposition, respectively, relative to the same WRF reference with an $8\times16$ decomposition. Lines indicate cross-case medians, and shading of the same colour spans the 25th--75th percentiles; symbols mark the measured metrics at 3, 6, 12, 24, 72, 120, 168, and 240\,h, with 27 cases at each lead time. The horizontal axis is logarithmic, and water vapour is expressed in g\,kg$^{-1}$.}
\label{fig:lead-mae}
\end{figure}

\begin{figure}[p]
\centering
\includegraphics[width=\textwidth]{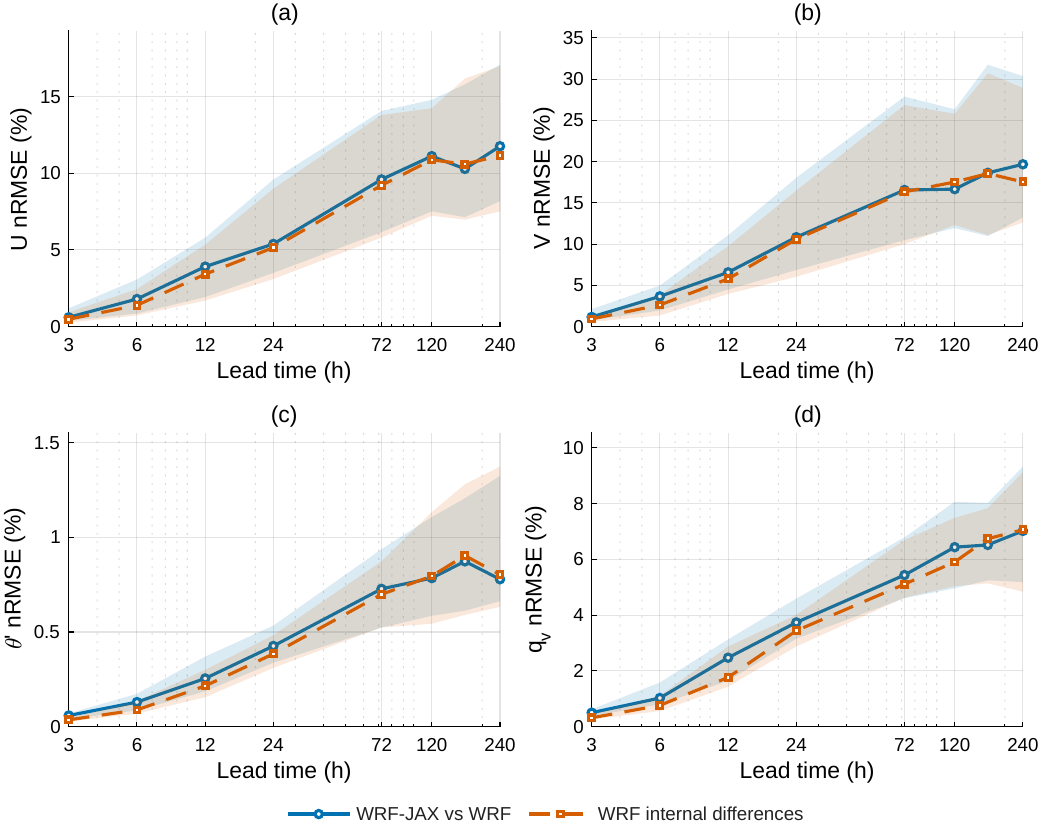}
\caption{As in Fig.~\ref{fig:lead-mae}, but for nRMSE, expressed in \%. Each case uses the root-mean-square amplitude of the reference field from the $8\times16$ decomposition at the same time as the normalization denominator, before the cross-case median and 25th--75th percentiles are calculated. All four variables have 27 valid cases at all eight analysis times; the potential-temperature metric refers to perturbation potential temperature $\theta'$.}
\label{fig:lead-nrmse}
\end{figure}

Figures~\ref{fig:lead-mae} and \ref{fig:lead-nrmse} supplement the evaluation with MAE and nRMSE evolution for all 27 cases, using cross-case medians and interquartile ranges consistent with Fig.~\ref{fig:lead}. MAE describes the mean absolute deviation in the original physical units; nRMSE is normalized separately for each case according to Eq.~(\ref{eq:nrmse}) before cross-case aggregation. Both comparisons use the corresponding WRF output with an $8\times16$ decomposition as the reference.

\clearpage
\noappendix

\codeavailability{The WRF-JAX source code is publicly available under the GNU Lesser General Public License v3.0 (LGPLv3) from the GitHub repository \url{https://github.com/Tongji-AOS/WRF-JAX}. The Zenodo DOI for the code archive of WRF-JAX v1.1.28 used in this study is \url{https://doi.org/10.5281/zenodo.23631990}~\citep{wrfjax2026}.}
\dataavailability{The NCEP GDAS/FNL data driving the experiments are available at \url{https://doi.org/10.5065/D65Q4T4Z}.}
\authorcontribution{Shijin Yuan supervised the research, acquired funding, and reviewed and edited the manuscript. Tianyu Shen developed the methodology and WRF-JAX code, conducted all experiments, performed validation and visualization, and drafted, reviewed, and edited the manuscript. Bin Mu supervised the research, managed the project, and acquired funding. Zhen Wei contributed to visualization and manuscript review and editing. Bo Qin and Guokun Dai contributed to funding acquisition.}
\competinginterests{The authors declare that they have no conflict of interest.}
\begin{acknowledgements}
We thank the WRF and WPS developers and the JAX maintainers. Coastlines in the figures are from the public-domain Natural Earth dataset.
\end{acknowledgements}
\financialsupport{This study was supported in part by the Meteorological Joint Funds of the National Natural Science Foundation of China under Grant U2542211, U2542212, and in part by the Original Exploration Project of the National Natural Science Foundation of China under Grant 42450163, and in part by the National Natural Science Foundation of China under Grant 42635009, 42405147.}
\bibliographystyle{copernicus}
\bibliography{references}
\end{document}